\documentclass[letterpaper]{sig-alternate-10pt}
\usepackage{cite}
\usepackage{times}
\usepackage{dsfont}

\usepackage{graphicx}
\usepackage{graphics}
\usepackage{subfigure}
\usepackage{epsfig}
\usepackage{latexsym}
\usepackage{amsfonts}
\usepackage{amssymb}
\usepackage{paralist}
\usepackage{comment}
\usepackage{xspace}
\usepackage{mathrsfs}
\usepackage{slashbox}
\usepackage{color}
\usepackage{cases}

\usepackage{url,epsfig,array}
\usepackage{leftidx}
\usepackage{amsmath}
\usepackage{paralist}
\usepackage[lined,linesnumbered]{algorithm2e}

\def\ie{\textit{i.e.}\xspace}
\def\etal{\textit{et al.}\xspace}

\def\eg{\textit{e.g.}\xspace}

\def\ourprotocol{\textbf{TEXIVE}\xspace}

\newcommand{\CUTIT}[1]{{}}

\begin{document}

\title{TEXIVE: Detecting Drivers Using Personal
Smart Phones by Leveraging Inertial Sensors\vspace{-0.2in}}
%\title{You're Driving and Texting!}
%\author{ACM MobiCom 2013, Paper ID: 12}
\author{Cheng Bo, Xuesi Jian, Xiang-Yang Li\\
Department of Computer Science, Illinois Institute of Technology,
Chicago IL\\
Email: cbo@hawk.iit.edu, xjian1@hawk.iit.edu, xli@cs.iit.edu\vspace{-0.2in}}

\maketitle

\begin{abstract}
In this work, we address a fundamental and critical task of detecting
the behavior of driving and texting  using smartphones carried by users.
We propose, design, and implement  \ourprotocol that
 leverages various sensors integrated in the smartphone and
realizes our goal of distinguishing drivers and passengers and
detecting texting using rich
user micro-movements and irregularities that can be detected by sensors
in the phone before and during driving and texting.
Without relying on external infrastructure, \ourprotocol has an
advantage of being readily implemented and adopted, while at the same
time raising a number of challenges that need to be carefully
addressed for achieving a successful detection with good
\emph{sensitivity}, \emph{specificity}, \emph{accuracy}, and
\emph{precision}.
Our system distinguishes the driver and passengers by  detecting
whether a user is entering a vehicle or not,  inferring
which side of the vehicle s/he is entering,  reasoning whether the
user is siting in front or rear seats, and discovering if a user is
texting by fusing multiple evidences
collected from accelerometer, magnetometer, and gyroscope sensors.
To validate our approach, we conduct extensive experiments
with several users on various vehicles and smartphones.
Our evaluation results show that \ourprotocol has a classification
accuracy of 87.18\%, and precision of 96.67\%.

\end{abstract}

%\begin{keywords}
%EDDIS, Inertial Sensor, Smartphone.
%\end{keywords}

\section{Introduction}
Distracted driving diverts driver's attention away from driving, which
will endanger the safety of driver, passenger, and even
pedestrian~\cite{distraction}.
One recent study indicates that every year, at lease $23\%$ of all
motor vehicle crashes or $1.3$ million crashes involve using cell
phones and texting~\cite{nhtsa}.
In United States, on a hand-held cell phone while driving is
considered illegal in $10$ states  and the District of
Columbia~\cite{iihs,mccartt2006cell}, and over $30$ states and District of Columbia
forbid texting message while driving.

The prevalence of cell phone and severe negative impact of driving and
texting on safety have stirred numerous researches and innovations on detecting
and preventing the behaviors of driving and texting.
The majority  effort has been on detecting this behavior using
various technologies, such as mounting a camera to
monitor  the driver  \cite{wahlstrom2003vision,herrmann2010hand},
relying on acoustic
ranging through car speakers \cite{yang2011detecting}, or leveraging
sensors  and cloud computing to determine driver
\cite{chu2011vehicle}.
Another line of innovations is to prevent driver from using
phones  \cite{shabeer2012unified} via signal jammer.
Recently, a number of apps have been developed to report driving and
texting, \eg, Rode Dog \cite{rode}.
These apps  could not distinguish between the driver and  passengers.
These techniques have been shown to perform well under various
circumstances and assumptions, but not without some limitations, \eg,
using extra
infrastructures such as cameras or radio interceptor
\cite{wahlstrom2003vision,herrmann2010hand,shabeer2012unified},
requiring special vehicles (Bluetooth and special layout of speakers
\cite{yang2011detecting}), or collaboration of multiple phones in the
vehicle and cloud computing \cite{chu2011vehicle}.

In this work, we address the fundamental and critical task of detecting
the behavior of driving and texting  by leveraging the smartphones
carried by users.
Our system leverages various sensors (\eg, accelerometer, gyroscope,
magnetometer sensor) integrated in the smartphone and
realizes our goal of distinguishing drivers and passengers and
detecting texting using rich
user micro-movements and irregularities that can be detected by these sensors
 \emph{before} and \emph{during} driving and texting.
Our system distinguishes the driver and passengers by  performing the
following tasks by fusing multiple evidences
collected from accelerometer, magnetometer, and gyroscope sensors: 1)detecting
whether a user is entering a vehicle or not,  2)inferring
which side of the vehicle s/he is entering,  3)reasoning whether the
user is siting in front or rear seats, and 4)discovering if a user is
texting.
Without relying on external infrastructure, our system has an
advantage of being readily implemented and adopted, while at the same
time raising a number of challenges that need to be carefully
addressed for achieving a successful detection with good
\emph{sensitivity}, \emph{specificity}, \emph{accuracy}, and
\emph{precision}.
A common challenge  is
to minimize or even remove the negative impact of the inherent noise
of the data collected by these sensors.
Another challenge is to increase the sensitivity, accuracy, and
 precision of detection, which is extremely difficult because of the
 potential pattern similarities among different user activities and
 the vast possibilities of how a user will drive and text (\eg, how a
 user will carry the phones, where the user will put the phone, how a
 user will enter the car, how will a  user sit in the car).
The third challenge is to design a real-time activity detection and
recognition with high accuracy and energy efficiency.

To detect whether a user starts entering a vehicle as a driver, or is doing some
other activities (such as walking, siting, or entering a public
transportation), we collected the data from accelerometer and
magnetometer when users are performing various activities and observed
some unique patterns (by converting the signal to the frequency domain
using DCT and wavelet).
To infer whether a user enters the vehicle from left side or right
side of the vehicle, or sits in front or rear seats, we exploit the
unique patterns in the accelerometer and magnetometer data observed from
respective actions and make cognitive decision based on machine
learning techniques.
Our system carefully exploits some unique external phenomena:
1) when vehicle engine is started, the data from magnetometer exhibits
different patterns when users sit in front or back seats;
2) the accelerometer data experiences different curves when the phone
is placed in the front seats and back seats when the vehicle passes
through a bump or pothole;
3) the accelerometer data showed different and distinguishable
patterns when user enters the vehicle from different sides even the
user has the phone in different pockets.
To validate our approaches, we conduct extensive experiments of our
system with several kinds of vehicles and smartphones.
Our evaluation results show that our approach had a classification
accuracy of 87.18\%, and precision of 96.67\%.
%\mynote{more experiments results here}

The rest of paper is organized as follows.
In Section~\ref{sec:related} we briefly review the exiting work on
distinguishing driver and passengers, and in general activity
detection and recognition using inertial sensors.
We  present the system design in
Section~\ref{sec:design} by discussing how we tackle several critical
 tasks for detecting texting by driver.
Section~\ref{sec:energy} presents the energy consumption strategy of \ourprotocol.
We report our extensive evaluation  in
Section~\ref{sec:evaluation} and conclude the paper in
Section~\ref{sec:conclusion}.

\section{State of the Art}

A number of  innovative systems have been proposed and developed in
the literature to distinguish between the driver and passenger, or
prevent the driver from using the cellphone.
The first line of work is to use some external devices to detect
whether the driver is distracted or whether the driver is using a
phone~\cite{yang2011detecting,sathyanarayana2008body,chu2011poster}.
Bergasa \etal \cite{Bergasa2006real} designed a feasible prototype  to
detect the driver distraction using special wearable equipment, a
circuit equipped with a camera to monitor the driver's vigilance in
real time.
Kutila \etal \cite{kutila2007driver} developed another smart
human-machine interface  to measure the driver's momentary
state by fusing stereo vision and lane tracking
data~\cite{kutila2007driver}.
Although these two systems could detect the driver's distraction, they
do not take the hand held devices into account, and the detection
accuracy is approximately $77\%$ on average.
Salvucci \cite{salvucci2001predicting} built a cognitive architecture
to predict the effects of in-car interface on driver's behavior based
on cell phone.
With the increasing number of accidents caused by using cell phone
while driving, many efforts focus on reducing the dangerous driving
distraction, but allowing drivers to deal with the devices with less
effort, such as Blind Sight~\cite{li2008blindsight},
Negotiator~\cite{wiberg2005managing}, Quiet
Calls~\cite{nelson2001quiet}, and Lindqvist~\cite{lindqvist2011undistracted}.
Most of the aforementioned designs require extra equipments or modifying
cars to assist  detecting the drivers' activity, which will
increase the system cost  and coordination difficulty,
 or  fail to take the presence of hand held smartphones
into account.

Other existing solutions for distinguishing driver and passenger
 rely on specific mechanisms to determine the location of the
 smartphones.
For example, recently, Yang \etal \cite{yang2011detecting} present an
innovative method by leveraging the high frequency beeps from
smartphone over Bluetooth connection through car's stereo system, and
calculate the relative delay between the signal from
speakers to estimate the location of
smartphone~\cite{yang2011detecting}.
However, a possible obstacle to this system is the requirement of
using Bluetooth, which may be not available in most old cars as well as  new models either.
Even with Bluetooth, because of the varying cabin sizes and stereo
configurations, the accuracy may be compromised in some extent.
Chu \etal \cite{chu2011poster}  presented a phone based sensing system
to determine if a user in a moving vehicle
is a driver or a passenger without relying on additional wearable sensors or custom
instrumentation in the vehicle.
They relied on collaboration of multiple phones to
process in-car noise and used a back-end cloud service in differentiating
the front seat driver from a back seat passenger.
Compared with these systems, our system will only use the smartphone
carried by the driver and does not require special devices in the car.

Our approaches involve a number of activity detection and recognition
using inertial sensors integrated in  smartphones, which has been
studied for various different purposes~\cite{kwapisz2011activity}.
Bao \etal \cite{bao2004activity} performed activity recognition
 based on multiple accelerometer sensors,  deployed on specific parts
 of human body, such as wrist, arm, ankle, or thigh.
Parkka \etal \cite{parkka2006activity} proposed a system
 embedded with  $20$ different wearable sensors to recognize
 activities.
Tapia \etal \cite{tapia2007real} presented a real-time algorithm to
 recognize not only physical activities, but also their intensities
 using five accelerometers and a wireless heart rate monitor.
Krishnan \etal ~\cite{krishnan2008real} demonstrated that putting
 accelerometers in certain parts is inadequate to identify activities,
 such as sitting, walking, or running.
Mannini \etal \cite{mannini2010machine} introduced multiple machine
learning methods to classify human body postures and activities,
including lying, sitting, running, and climbing, based on
accelerometers attached to certain positions on the body.
Lee \etal \cite{lee2002activity} introduced a novel system to identify
 user's location and activities through accelerometer and angular
 velocity sensor in the pocket, combined with a compass on the waist.
Ravi \etal \cite{ravi2005activity} used HP iPAQ to collect
acceleration data from sensors wirelessly, and recognize the motion
activities.
A few studies perform activity detection and recognition using
commercial mobile
devices~\cite{kwapisz2011activity,miluzzo2008sensing,yang2009toward,brezmes2009activity},
which are more practical and  unobtrusive.
Unfortunately, these systems and approaches cannot be used for
distinguishing driver
and passenger and detecting driving and texting activities.

%In this work we focus on activity recognition and detection related to
%driving and texting.

\section{System Design}
To address the driver-passenger challenge, we will leverage the
existing inertial sensors integrated in  smartphones and exploit some unique
and distinguishable patterns observed from a sequence of
sensory data.
In this section, we discuss in detail design goals, the approach to
detect which side a user is entering the car, and which row the user
is sitting, for location classification.

\subsection{Challenges and Design Goals}
%%% this subsection will discuss the challenges and goals

The purpose of our system is to distinguish the driver and the passengers
 using smartphone only
 without any assistance of dedicated infrastructures or intrusive
 equipments in the vehicle.
%Our system will adopt a non-intrusive and pure smartphone based
% solution.
The key goal that led to our inertial sensor approach is to be able
 to determine various activities and classify the phone location from
 the observed unique and distinguishable micro-movements and
 sequential patterns.
This pure phone software solution, however, leads to several technical
 challenges:

 \textbf{Robust to Diversity and Ambiguity of Human Activities:}
The system requires a real-time activity recognition to identify the
 driver with high accuracy and low  delay.
For example, when we know that a user is walking towards the car, we
 should start the algorithm to determine if s/he enters the car or
 not, and  determine if s/he is a driver or  passenger.
However, because of the difference of smartphone's orientation,
 position and location, same user activity may result multiple
 irregular sensory data.
In addition, we need an effective method to detect the starting point
 and the duration of an activity for the purpose of increasing the
 accuracy while considering the randomness of action even by the same
 user.
We assume that the behaviors between drivers
  and passengers are different, especially during the action of
  entering the vehicle and driving.
We need to carefully identify the signal and patterns that can be used
 for accurate distinguishing.

 \textbf{Robust to Data Noise:} It is widely accepted and
  verified in our testing that the  data collected by the inertial
  sensors in smartphones contain significant noise.
If it is not carefully addressed, such noise may override the small
 changes caused by different human activities.

 \textbf{Computation and Energy Efficiency:}
 As smartphones have limited computation capability and limited energy
 supply, standard smartphone platforms should be able to execute the
 system in an online manner with running time of seconds or
 sub-second.
The system will be running at background in carefully selected dynamic
 duty cycle.
Thus we need a careful tradeoffs between efficiency and detection accuracy.

\subsection{System Architecture Overview}
%%%% this subsection will present an overview of the system

\begin{figure}[hptb]
\vspace{-0.1in}
\centering
\includegraphics[height=1.75in,width=2.75in]{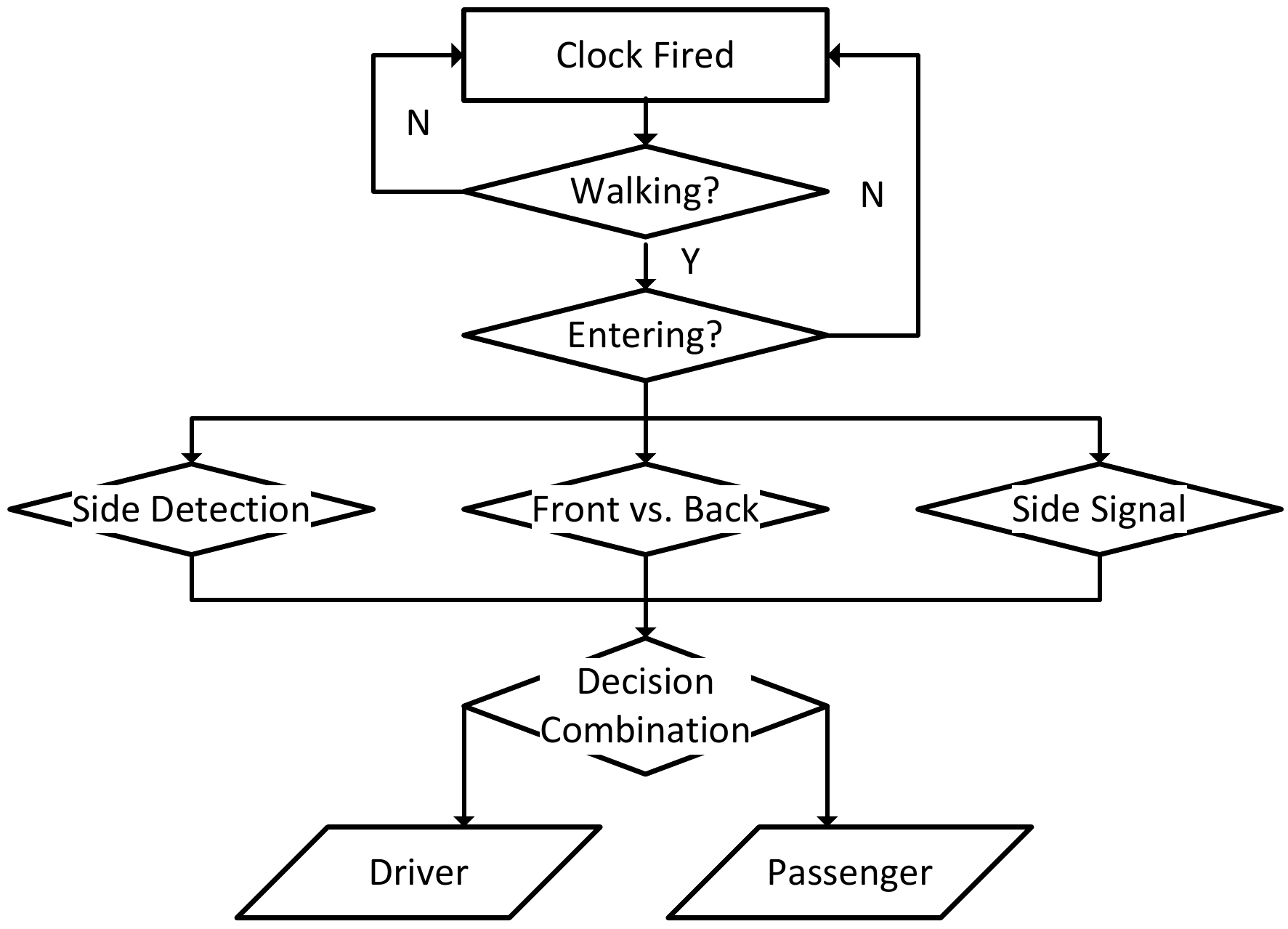}
\vspace{-0.1in}
\caption{The system overview}% here P is passenger and D is driver.}
\label{fig:process}
\vspace{-0.25in}
\end{figure}

We now briefly describe the architecture of our system.
%The purpose of the system is to use the smartphone to recognize the
%user behavior, locate the position of the smartphone in a vehicle, and
%identify whether the user is a driver or a passenger without any
%additional intrusive equipment.
In this work we propose a three-phase solution to accomplish the task:
  recognizing the action,  locating the user in the
 vehicle, and  determining the role of the user based on assembled
 information.
Figure~\ref{fig:process} illustrates various components of our system
that will be described in details.

\textbf{Activity Identification:}
When driving or sitting in the vehicle, the behaviors are different
 from most of the other activities in our daily lives.
Empirically, such driving activities usually start from walking
 towards the vehicle, and are followed by entering the vehicle,
 fastening the seat belt, or even pressing the gear and brake.
Thus, walking will be detected periodically, and we looking for the action of
 entering the vehicle following the detection of the walking action.
Notice that our system does \textbf{not} require \emph{any}
 interaction from the user, thus, it is critical to find when a user
 will enter the vehicle so that no driving and texting activity is
 missed.
Generally, most of users get used to carry smartphone for a whole day,
 the system will definitely record multiple kinds of behaviors throughout
 one day.
A research task here is to identify related activities from a rich set
 of potential daily activities,  including walking, sitting, standing or going
upstairs.
We observe that most time different activities will be reflected in
different micro-movements and motion patterns, although sometimes
different activities will have similar ``observed'' patterns.
We will exploit some subtle differences (\eg, the different frequency
distribution when converting the time-series data to the frequency
domain, the variance of the amplitude of the time-series data) between
observed data from various sensors (\eg, accelerometer,  magnetometer,
 and gyroscope) for
 recognizing driving activity from other activities.

By collecting the daily activities, we study the distribution of
 activities and temporal and spatial relationship between different
 activities, and construct a Hidden Markov Model
 (HMM)~\cite{rabiner1986introduction} to analyze the behavior based on
 the observed sensory data.
This model will then be used to further reduce the energy consumption by
carefully adjusting the operating duty-cycle of the system.

\textbf{Detecting Left vs. Right:}
The second component of our system is to determine whether a user
entered the vehicle from the left side of the vehicle or the right
side of the vehicle.
If the user is recognized to have entered the vehicle from the right
 side, taking US as an example, the user must be a passenger usually.
But we still cannot judge the role precisely if the user is from the left side.
We found that the accelerometer data exhibits different patterns when
a user entered the vehicle from different sides and having the smartphone
at different pockets.

\textbf{Detecting Front vs. Back:}
Detecting the side cannot  uniquely identify the driver.
Thus, the third phase is proposed to determine whether the user is
sitting in the front seats or back seats.
Together with the side information, we can determine the location of
 the phone in a vehicle.
Suppose there are one driver seat and three passenger seats  in the
 vehicle.
Take US as an example, the user must be the driver if he is sitting in
 the left side of the front row.
Our approach relies on two separate signals.
The \emph{first} signal is the change of magnetic field strength value.
 Our extensive tests show that when the phone  is in the  front half
 of the car, we can see an obvious change in the collected
 magnetometer data when the vehicle engine is started.
The \emph{second} signal is the change of the accelerometer value based on
different road condition.
We observed that when a car passing through a bump or a pothole,
 there are unique and distinguishable patterns when the phone is in
 the front seats or the back seats.
The bump signal, although not guaranteed to happen, can always
 accurately  determine whether the phone is in front seats or rear seats.

\textbf{Further Improvement:}
Although these phases provide us some information of the behavior and
 location of the user in a vehicle, we cannot neglect the issue
 regarding the identification accuracy.
In this work we rely on machine learning process and evidence fusion
to further improve the accuracy.
For example, when driving, the driver may text in an abnormal patterns,
 while the passenger, on the other hand, may still follow regular patterns.

%To further improve the robustness of our system,
% we consider the diversity of human activities, \eg,
% considering the fact that users put their smartphone in different
% locations according to individual habit~\cite{ichikawa2005s}: $34\%$
% of  users put their smartphones in the trousers pocket, while only
% $5\%$ put in the upper body pocket.
To further improve the robustness of our system,
 we consider the diversity of human activities, \eg,
 considering the fact that users put their smartphone in different
 locations according to individual habit~\cite{ichikawa2005s}.

Another issue worth mentioning is that the system  will be running at
background, and operating the entering-vehicle  recognition and side
detection in real time, rather than keep recording the sensory data
into local buffer and detect the activities through rolling back which
is most common way.
The strategy is determined based on two main reasons, efficiency and
reducing cost.

\begin{figure}[hptb]
\centering
\vspace{-0.1in}
\includegraphics[scale=0.5]{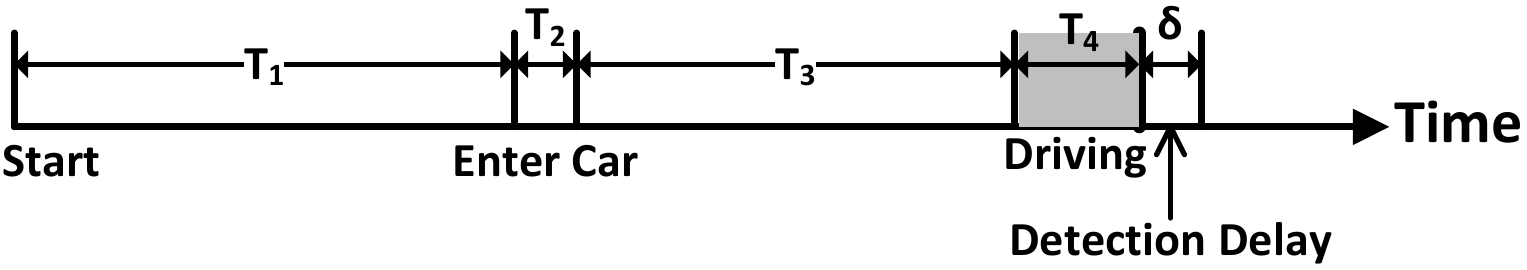}
\vspace{-0.15in}
\caption{Real time strategy vs. rolling-back.}
\label{fig:strategy1}
\vspace{-0.15in}
\end{figure}

Suppose the  system starts recording the sensory data at time $0$, as shown in
Figure~\ref{fig:strategy1}.
At time  $T_1$, the user starts entering a car which lasts
 $T_2$.
He starts driving after a delay of $T_3$ sitting inside the car.
It is common that a user may make a phone call before driving.
Once the system detects the driving behavior with detection delay
 $\delta$, after users has driven for time $T_4$.
The whole duration of  the sequence of
 actions will last $T_1 + T_2 + T_3 + T_4 + \delta$.
However, the exact duration of every $T_i$ is unknown and
 unpredictable, the amount of sensory data which have to be stored in
 the buffer  will be extremely large if we do offline detection.
While in our real-time detection system \ourprotocol,
 we can distinguish driver at time $T_1+T_2$ without buffering data.

\subsection{Inertial Sensors and Data Noise}

Suppose the smartphone is carried by the user and placed in a
pocket, the motion of a human body is reflected on the motion of
smartphone through three inertial sensors (accelerometer,
magnetometer, and gyroscope).
In our system, the sampling rate is set as $20Hz$, and then a series
of changing values on the sensor readings will represent the
continuous human activity.

As a rigid body,
the readings from sensors only apply to the coordinate of smartphone,
which could be denoted as Body Frame Coordinate (BFC).
Since the motion condition of smartphone is irregular and unpredictable,
 without knowing the posture of the phone in the pocket, it
is difficult to analyze the human behavior in detail.
On the other hand, from the perspective of the earth,
individual person may act by following some hidden regular pattern
unconsciously, and the only difference may be the frequency, duration
and the amplitude of the behavior.
In this case, we extract the readings from the sensors, and convert
the value into the Earth Frame Coordinate (EFC) to represent the activity.

%The accelerometer senses  both the linear acceleration and gravity
%applied on the three axes in BFC, and the magnetometer monitors the
%changes in the ambient magnetic field at the same coordinate system.
%Generally, the raw magnetic field value reflects the earth magnetic
% field under the current orientation of the sensor.
%Thus, the combination of gravity and magnetic field will lead to the
% orientation of the smartphone from the perspective of earth.

In addition, in order to reduce the noise coming from both intrinsic mechanical
 structure and measurement, we adopt a calibrated orientation method
 through Extended Kalman Filter (EKF).
%After obtaining the initial Euler angle, the sensor fusion process is
% followed.
%Obviously, the linear acceleration from the perspective of Earth is
%enabled through rotating the linear acceleration from BFC to EFC
%according to continuous orientation.

\subsection{Entering Vehicles?}
%%% this subsection will discuss methods for decting whether a user
%%% will enter a vehicle: if enter, then starting the system, saving
%%% energy

A key challenge of this system is to identify specific activities in
real-time, especially determining whether a user will enter a vehicle
or is just performing other activities, which have similar observable
patterns as that of entering a vehicle.
Here we will mainly focus on the analysis of pattern recognition,
 including the walking and entering the vehicle.
The system is running at the background and will capture sensory data
 from three inertial sensors according to sophisticated duty cycle
 strategy.
Then, the acquired data will be processed and classified into
 specific activities through specially designed activity recognition
 method.

\paragraph{Activity Recognition}
\begin{figure*}[!ht]
\centering
\subfigure[ in trouser pockets\label{fig:walk_jean}]
{\includegraphics[scale = 0.25]{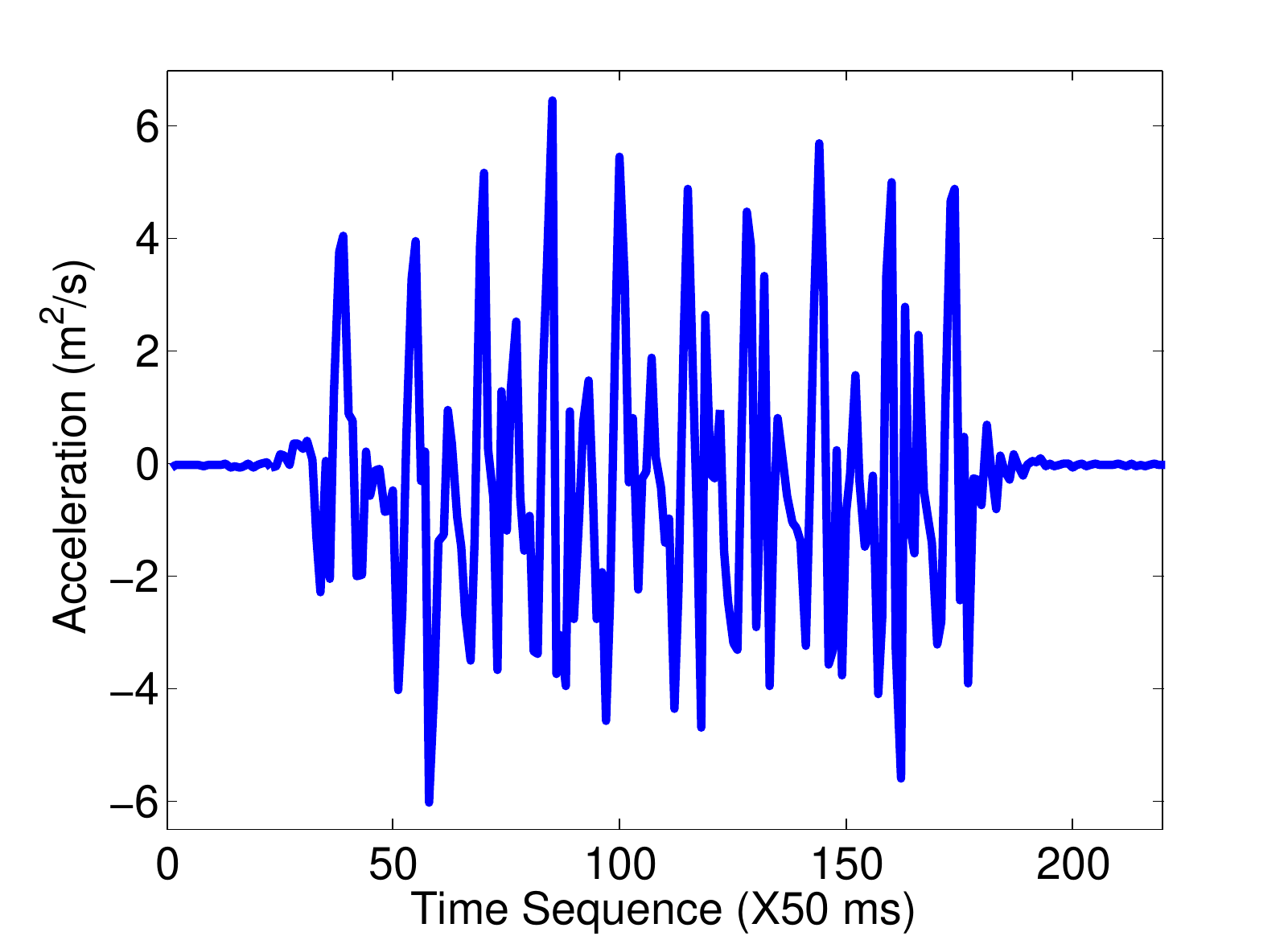}}
%\subfigure[ in shirt pockets\label{fig:walk_shirt}]
%{\includegraphics[scale = 0.2]{figures/walk_shirt.pdf}}
\subfigure[ascending stairs\label{fig:stairs}]
{\includegraphics[scale = 0.25]{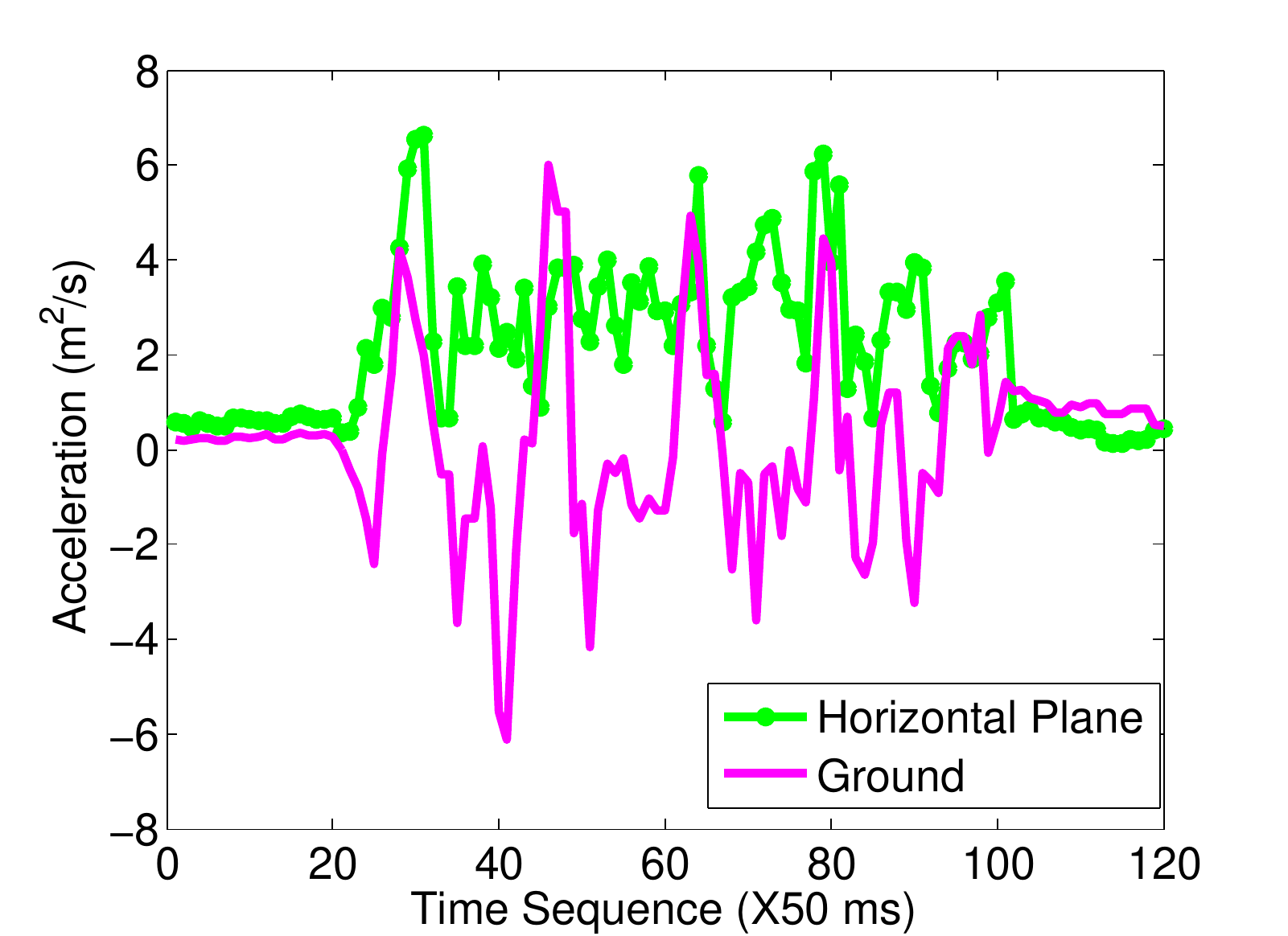}}
\subfigure[sitting down\label{fig:sit}]
{\includegraphics[scale = 0.25]{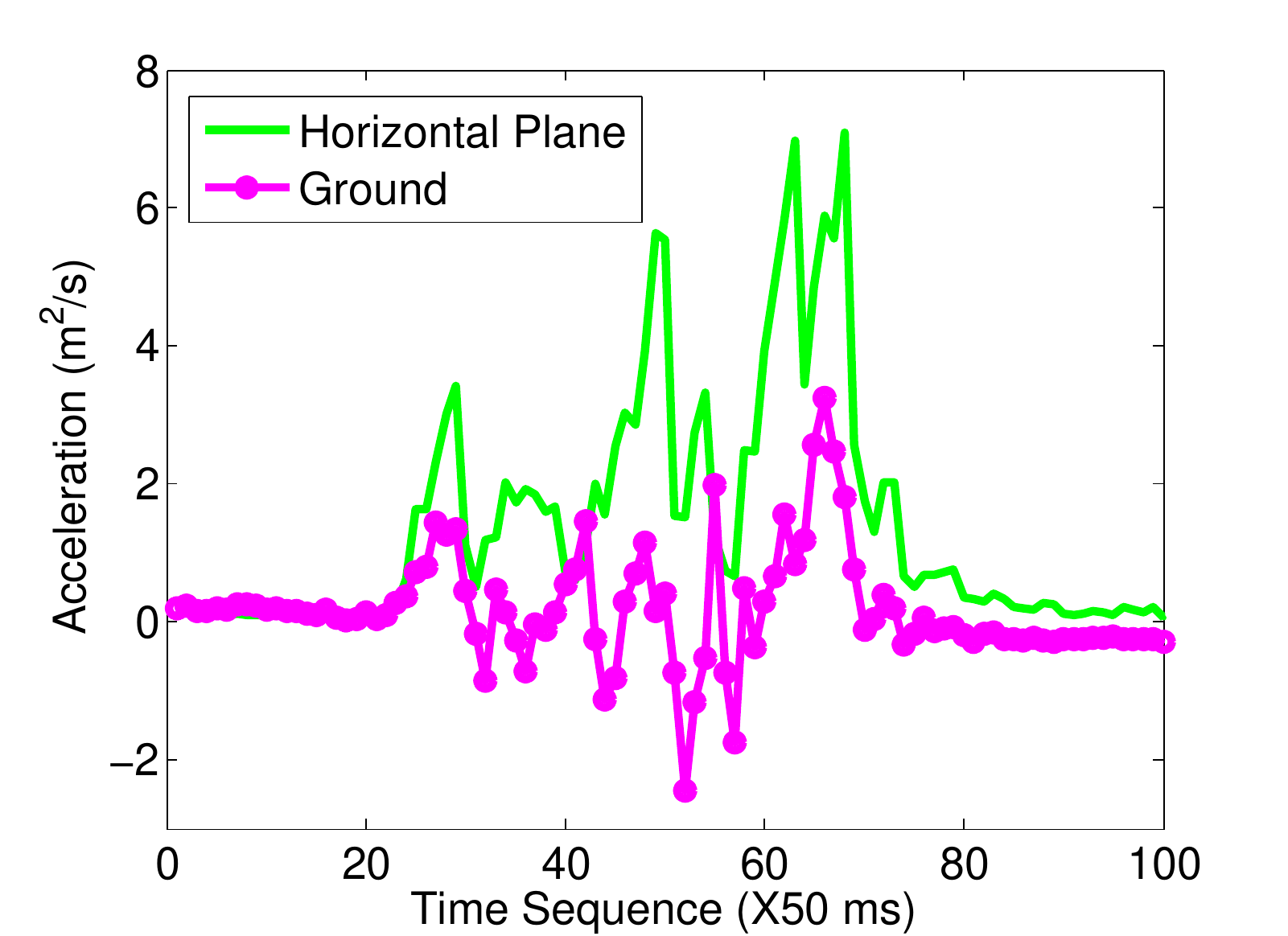}}
\subfigure[getting on the bus\label{fig:get_on_bus}]
{\includegraphics[scale = 0.25]{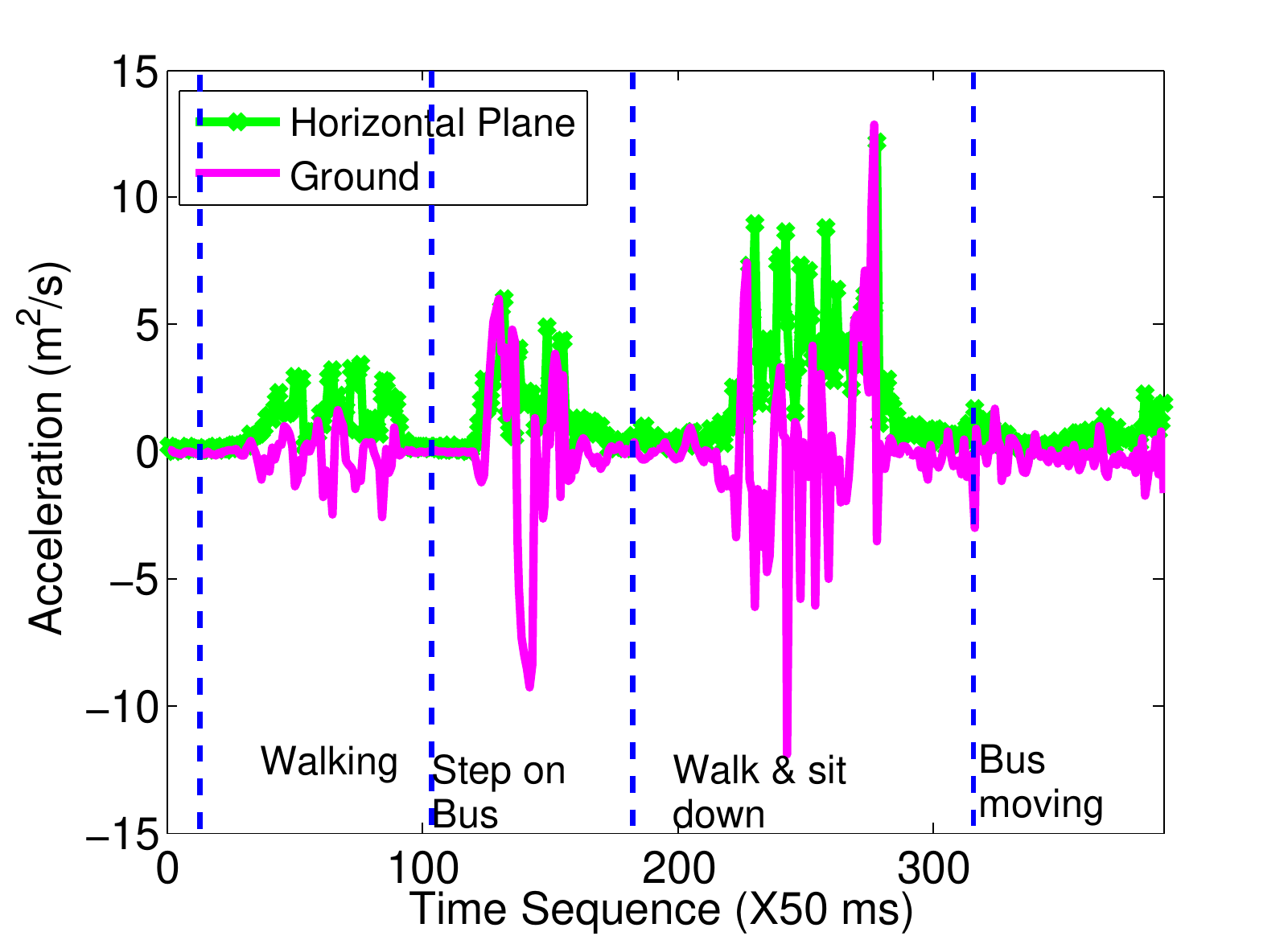}}
\vspace{-0.15in}
\caption{Sensory data extracted from accelerometer in different
  behaviors.}
\label{fig:walking}
\vspace{-0.21in}
\end{figure*}
As the sensors will collect data continuously, a critical step in
 activity recognition is to identify the starting time and ending time
 of an activity from a sequence of temporal data collected.
The second critical step in activity recognition is an effective
 feature extraction, which will  be the foundation for machine
 learning based activity recognition.
In our system, we adopt Discrete Cosine Transform
 (DCT)~\cite{ahmed1974discrete} to extract features from linear
 acceleration to represent the specific action.
Although the system mainly focuses on detecting driving activity, the
 system will  also encounter many other  activities as it is
 running in the background with a carefully selected adaptive
 duty-cycle.
For the purpose of driving activity detection and establishing
 HMM, we classify activities
 into three categories \emph{walking}, \emph{entering vehicle}, and
 \emph{other activities} (including sitting, going upstairs, downstairs, or
 getting on the bus).
%After we get the orientation and the linear acceleration from the
% perspective of Earth in every sample point, the observation of the
% activity could be reflected in time domain.

Both the walking behavior and going upstairs or downstairs involve repetitive motions,
 and the activity pattern could be reflected from the acceleration on the direction of
 ground, as shown in Figure~\ref{fig:walk_jean}and~\ref{fig:stairs}.
Sitting down is another activity which will be detected multiple times throughout
 a day, the pattern is illustrated in Figure~\ref{fig:sit}.
The behavior of getting on the bus is more complicated than the rest, because it
 consists of multiple other activities and the duration is much longer than the others,
 as shown in Figure~\ref{fig:get_on_bus}.
However, the patterns of these behaviors are different from each other, and it is not
 that difficult to distinguish each other.
In order to evaluate the performance of the activity recognition, we
 monitor the behavior for one specific user for one week, and collect
 multiple cases of sensory data.
In our initial test, we collect the activities of walking, sitting down, and going upstairs
 $20$ cases each, and $100$ other behaviors, including running, jumping, jogging.
In this work, we choose naive Bayesian
 classifiers~\cite{langley1992analysis} to detect and identify
 activities related to driving.
Naive Bayesian classifier could distinguish activities from the other
 in an acceptable accuracy ($91.25\%$).

Our system is carefully designed to  meet the requirement of online
 learning and real-time classifying.
It  collects the inertial sensory data while triggers the signal
 when specific activities are detected.
It also adjusts the activity model in real-time as new training
 examples from users are collected.
The protocol will adapt the new feature changes over time, train and
 reconstruct the model as the system being applied to other user.

\paragraph{Entering Vehicles?}
We first extract the feature of entering the vehicle by conducting
 extensive testing.
Typically, the activity of getting into the
 vehicle consists of following steps: walking towards the vehicle, opening
 the door, turning the body, entering, and sitting down.
Empirically, the duration of entering vehicle activity is relatively small.
In our system, we set the window size of the sampled data for activity
 recognition as $4.5$ seconds, which is based on the extensive
 evaluation to be presented later.
We then extract the feature regarding the linear acceleration in both the
 horizontal plane and ground direction in EFC.

In addition, the behavior will consist of two different cases
 according to the entering side, and such activity patterns are
 different.
Although the user's behavior could be reflected through build-in
 inertial sensors in attached smartphone, we cannot neglect the
 position where smartphones are put.
A recent study~\cite{ichikawa2005s} indicates that male users usually
 carry their phones in trouser pockets in most cases ($57\%$) while only $5\%$
 put in the upper body pocket.
We first study  the circumstance that the phone is in the pocket of trousers.

\begin{figure}[htbp]
\vspace{-0.1in}
\centering
\subfigure[Driver Side, Left Pocket\label{fig:ll2d}]{\includegraphics[scale = 0.25]{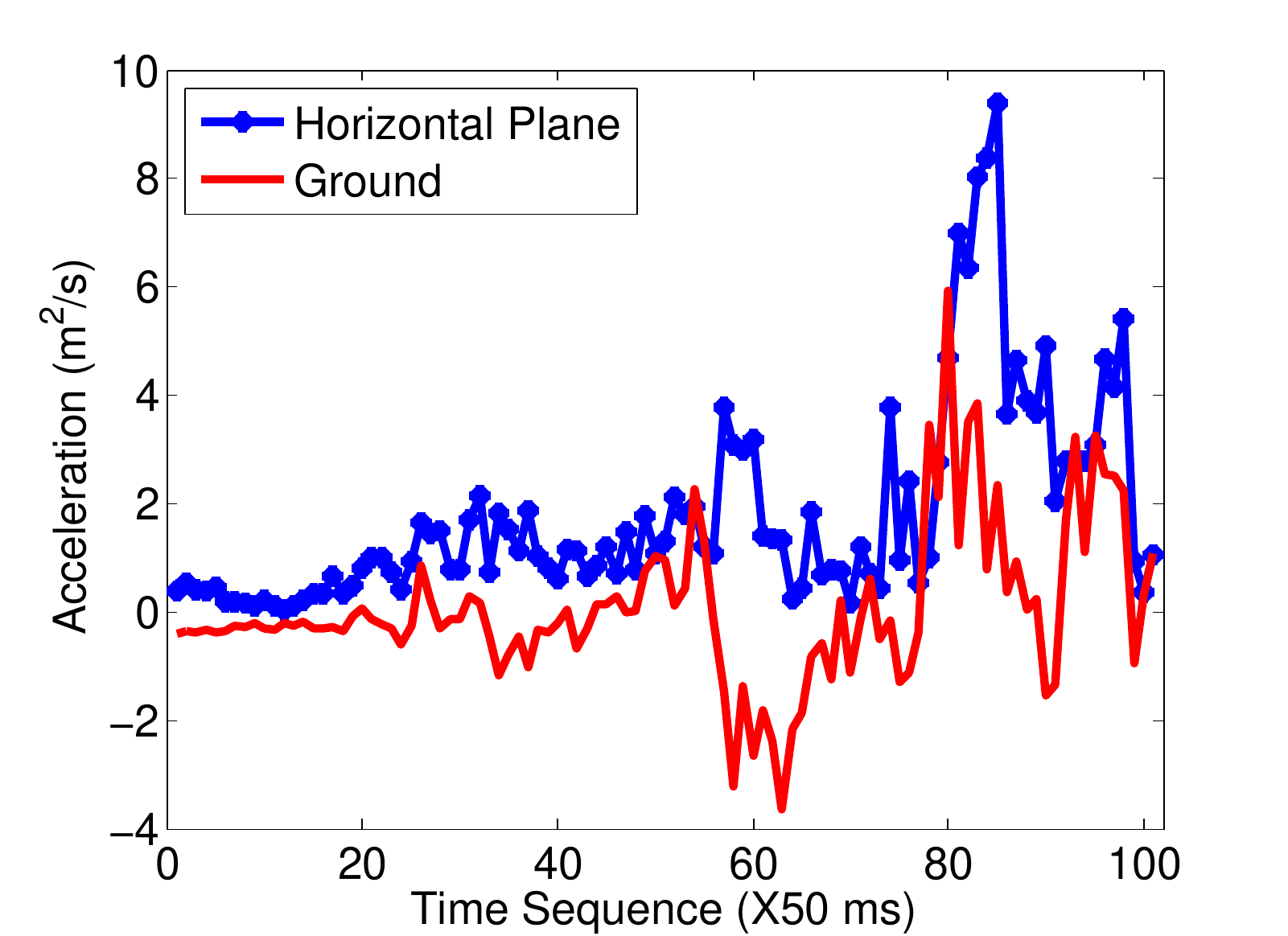}}
\subfigure[Driver Side, Right Pocket\label{fig:lr2d}]{\includegraphics[scale = 0.25]{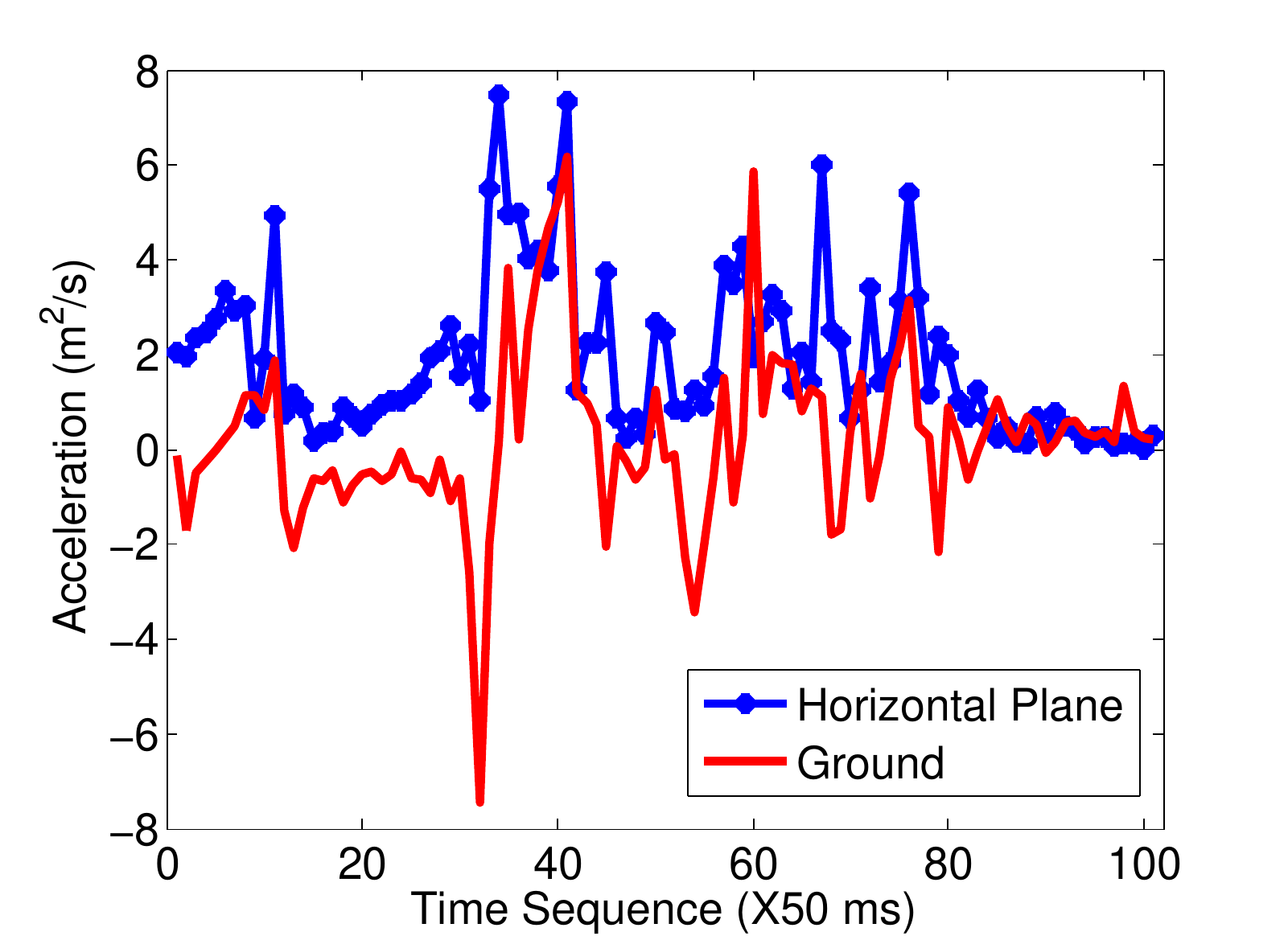}}
%\subfigure[Passenger Side, Left Pocket\label{fig:rl_enter}]{\includegraphics[scale = 0.25]{figures/rl2d.pdf}}
%\subfigure[Passenger Side,Right
%  Pocket\label{fig:rr_enter}]{\includegraphics[scale =
%    0.25]{figures/rr2d.pdf}}
\vspace{-0.1in}
\caption{Data extracted from accelerometer in Horizon plane and
  ground when people entering vehicle.}
\label{fig:enter_2d}
\vspace{-0.2in}
\end{figure}

We take a set of testing of entering the vehicle from both sides in
 the parking lot by a group of colleagues with smartphones under left
 and right trousers  pocket respectively.
We collect $200$ samples of the entering-vehicle activity  from both
 sides.
Due to irregular and unpredictable orientations of the smartphone,
 we transform all the data from BFC to EFC.
% as shown in Figure~\ref{fig:enter_efc}.
Since in EFC, the \emph{X} and \emph{Y} axes record the data along the Magnetic North and
 East, with different orientation of the vehicle, the acceleration in time domain will also
 be different, which will lead to the mismatch in the following activity
 recognition process.
%Since the acceleration in EFC reflected by the three directions
% only, and with different initial orientation of the vehicle, the
% direction of getting on will be different.
%Since in EFC, the first two axes record the data in Magnetic North and
% East, with different orientation, the value in time domain will also be
% different, which will lead to the mismatch in the following activity
% recognition process.
On the other hand, no matter which orientation the vehicle directs,
 the activity, from the perspective of the head of vehicle, will still
 be the same.
Thus, we calculate the vector of joint linear acceleration in
 the horizontal plane by $\sqrt{a_{north}^2 + a_{east}^2}$, and present
 the activity of entering the vehicle in both horizontal plane and ground
 direction in two cases in Figure~\ref{fig:enter_2d}, where the difference is much more obvious.

\paragraph{Improve the Accuracy}
Our initial test consists of $40$ cases of behavior of entering vehicle in total,
 and nearly $300$ hundred other behaviors, including walking, sitting down, ascending stairs,
 descending stairs, and jumping.
According to Naive Bayesian classifier, both the accuracy and precision
 of distinguishing entering vehicle from other actions are $84.46\%$ and $45.24\%$ respectively,
 as shown in Table~\ref{table:res1}.
\begin{table}[hptb]
\caption{Preliminary results on activity recognition.}
%\label{table:res1}
\begin{center}
\begin{tabular}{l|l|l}
\hline
   & Entering Vehicle & Other activities\\
\hline
  Test True & 38 & 46\\
  Test False & 3 & 250\\
\hline
\end{tabular}
\end{center}
\label{table:res1}
\vspace{-0.3in}
\end{table}
From the first experiment, although the behavior of entering the
 vehicle is easily identified through acceleration,
 a considerable number of other behaviors (sitting down mostly)
 are also identified as the same activity (the false positive is
 relatively high), which will  hinder the performance of the
 detection.
We discover that there are two main reasons for confusion sitting
 down with entering vehicle, for one the behaviors are quite similar
 some times (both may consist of walking, turning and sitting), and the
 other is that even if the same behavior may result in multiple
 patterns.
In order to overcome such phenomena, we propose a more comprehensive
 filter to elevate the result accuracy.

We observe that the main difference between regular sitting down and
 entering the vehicle is the environment:  the former  more likely
   happens indoor and  the latter is around a vehicle.
Such difference on environment happens to provide a key factor to
 distinguish the scenario, which is the amplitude of magnetic field.
\begin{figure}[hptb]
\centering
\vspace{-0.1in}
\subfigure[Toward a vehicle and
  sit\label{fig:start_car}]{\includegraphics[scale =
    0.25]{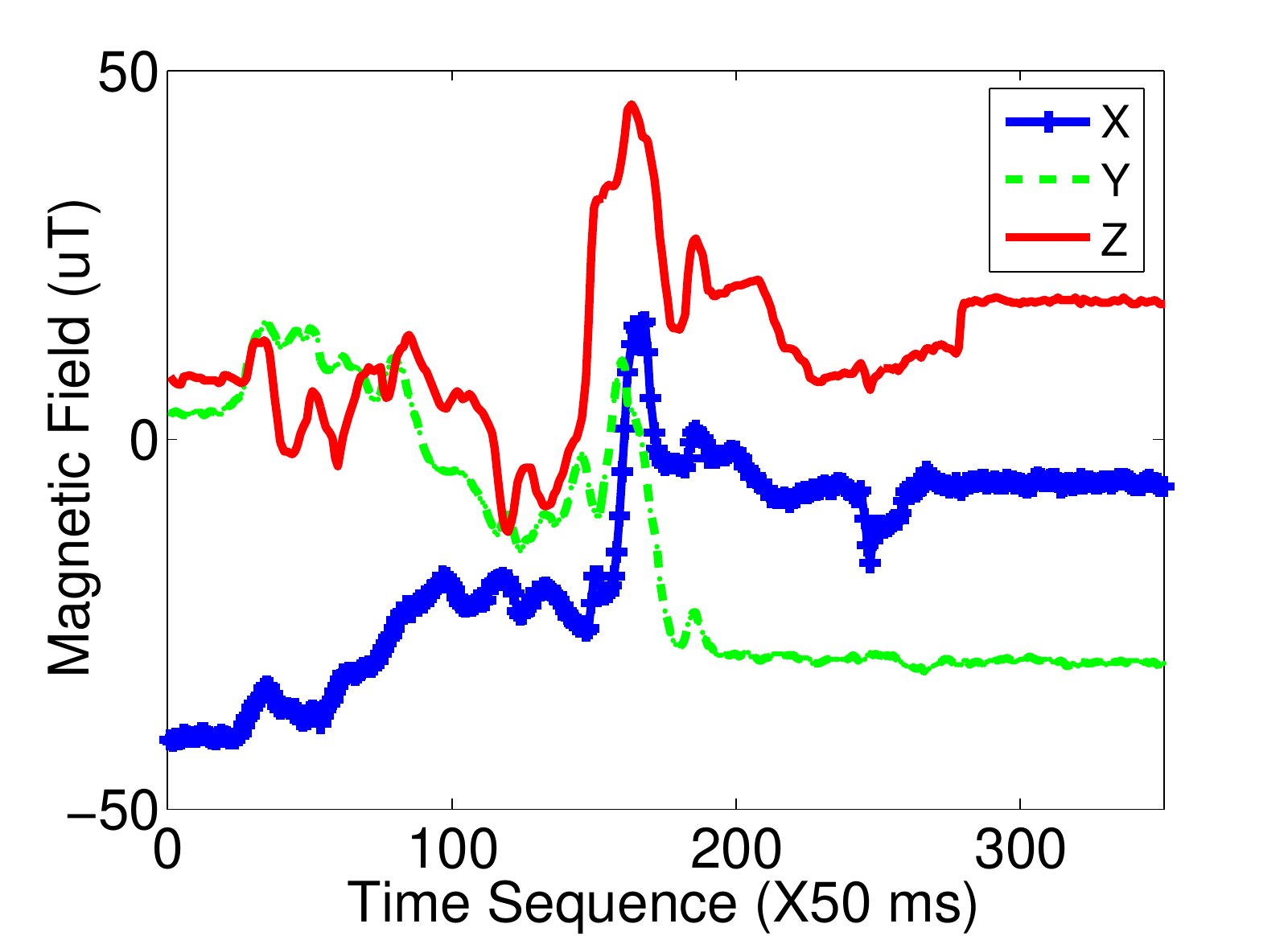}}
\subfigure[Toward a chair and
  sit\label{fig:mag_sit}]{\includegraphics[scale =
    0.25]{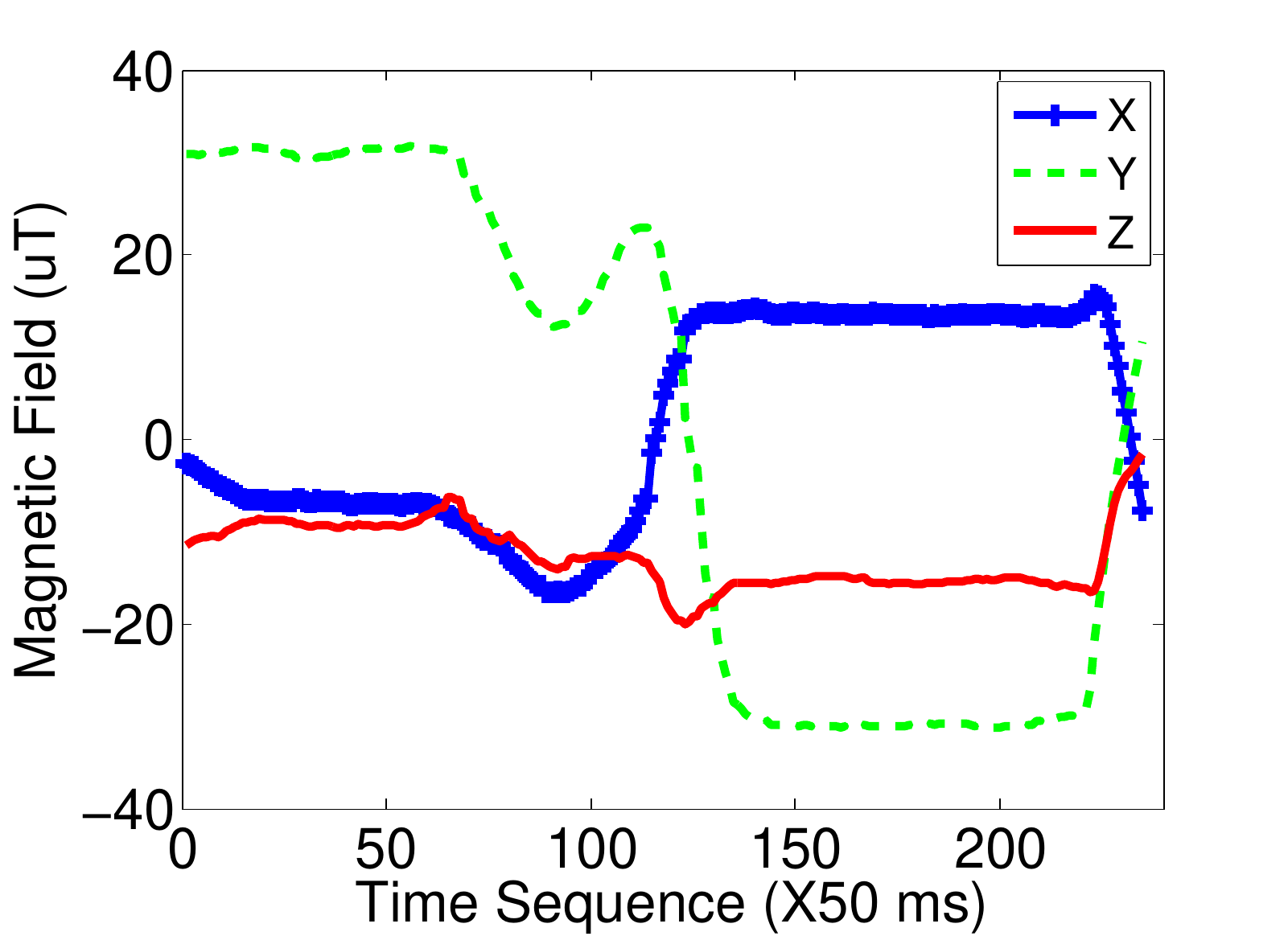}}
\vspace{-0.1in}
\caption{The magnetic field in two scenarios.}
\label{fig:mag_sit_car}
\vspace{-0.1in}
\end{figure}
Empirically, when approaching the vehicle, the magnetic field will
 vary dramatically because of the special material of vehicle and
 engine.
We sample the magnetic field data in three axes for both scenarios,
 and plot the data in Figure~\ref{fig:mag_sit_car}.
As we could see from the figures, the magnetic field fluctuates as
 user is approaching the vehicle, and becomes relatively stable when
 the user sitting in the vehicle.
When it comes to the other case, the magnetic field remains relatively
 stable even if the whole action contains walking and sitting down.
Besides, the accelerometer will detect large acceleration when
 the vehicle starts moving, which could be considered as a supplementary filter.
Such method could guarantee the user is in a vehicle.
%In this case, we take the pattern of magnetic field as a special
% filter to distinguish between sitting and entering-vehicle activity, and employ
% acceleration filter as a supplementary.
In our system, as long as the sitting action is detected, both the magnetic
 filter and acceleration filter will be triggered to scan the changing
 condition of ambient environment, and judge the current scenario.

\subsection{Left or Right Side?}
%%%% this subsection will present our methods for detecting whether a
%%%% user entering the vehicle from left or right side using the data
%%%% from accelerometer, using DCT and wavelet

The successful detection of entering the vehicle is an initial stage
 leads to the final driver detection.
However, such stage only constrains the location of the user in the
vehicle, but we still could not tell on which seat  the user is
sitting.
Then the second stage is to determine which side the user has entered the
vehicle as long as the entering behavior is identified, and we denote
this step as \emph{side detection}.

The side detection is based on the observation that the smartphone
 will experience a different movement when getting on the vehicle from
 driver side compared with that from passenger side.
Although the previous stage has already trained the behavior of
 getting on vehicle from both sides, the feature extraction according to
 DCT may consider some cases to be quite similar.
In our training stage, we test the behavior of entering vehicle in
 four different cases with respect to the location of smartphone and
 side of getting in.
Suppose the user is getting in the vehicle from driver side with his
 smartphone in the right trouser pocket, the motion will lead to a
 large fluctuation on acceleration and a more small one because of the
 inner leg entry followed by the other.
Here we denote the leg which is close to the vehicle and with smartphone in
 that pocket as the inner leg.
However, looking in to the case of getting on from passenger side with
 smartphone placed in the left trouser pocket, the observation is much
 the same.
%When transform the behavior according to DCT, the features are
%similar.
%The same case happens when it comes to the behavior of getting in the
%vehicle from driver side with smartphone in the left trousers pocket,
%and from passenger side with smartphone in the right trousers pocket.
The same thing happens when driver-side-left-pocket case versus right-side-right-pocket
case.

\begin{figure}[hptb]
\vspace{-0.1in}
\centering
\subfigure[Left Pocket, Driver Side\label{fig:ll_enter}]{\includegraphics[scale = 0.25]{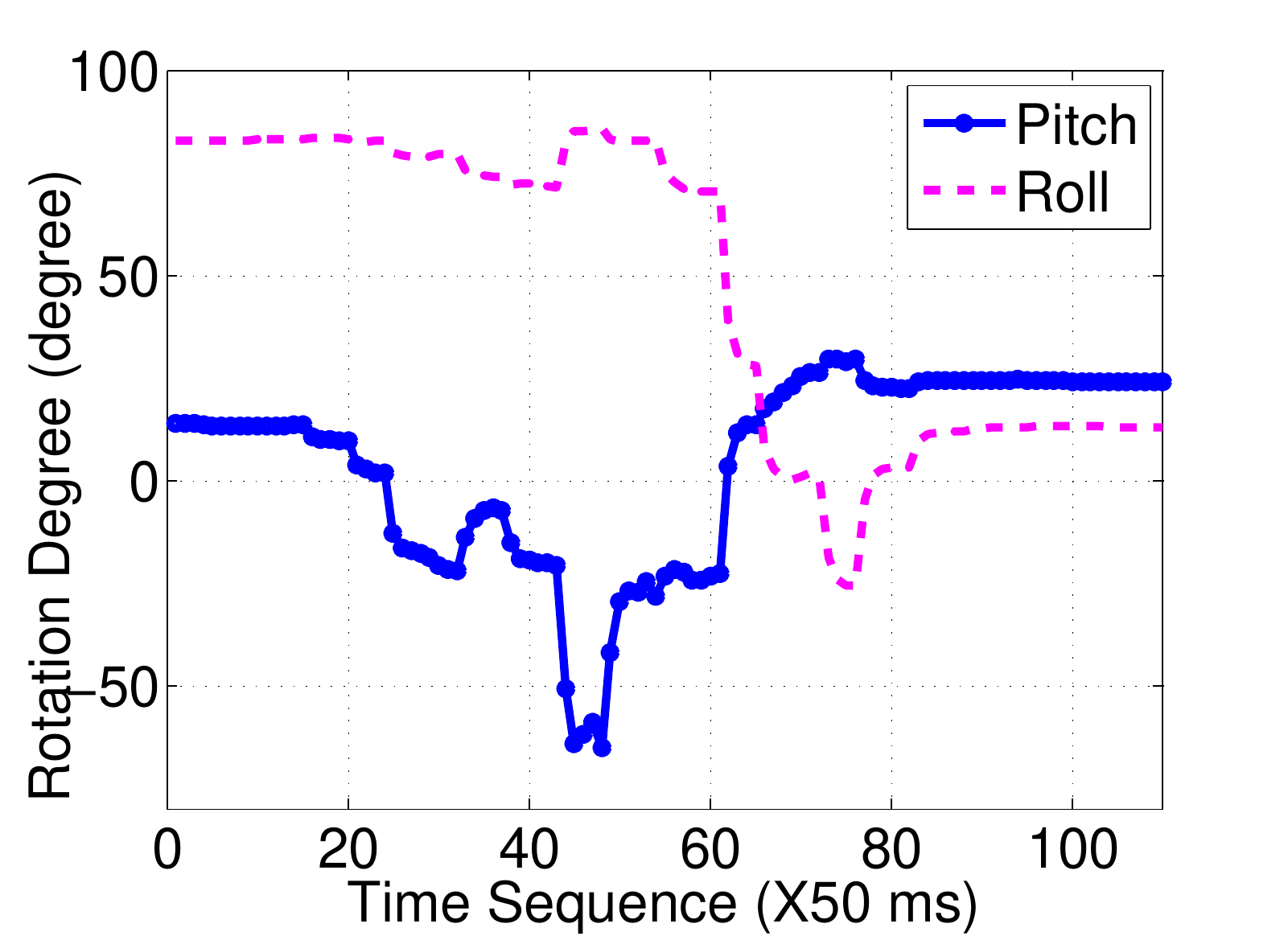}}
\subfigure[Right Pocket, Driver Side\label{fig:lr_enter}]{\includegraphics[scale = 0.25]{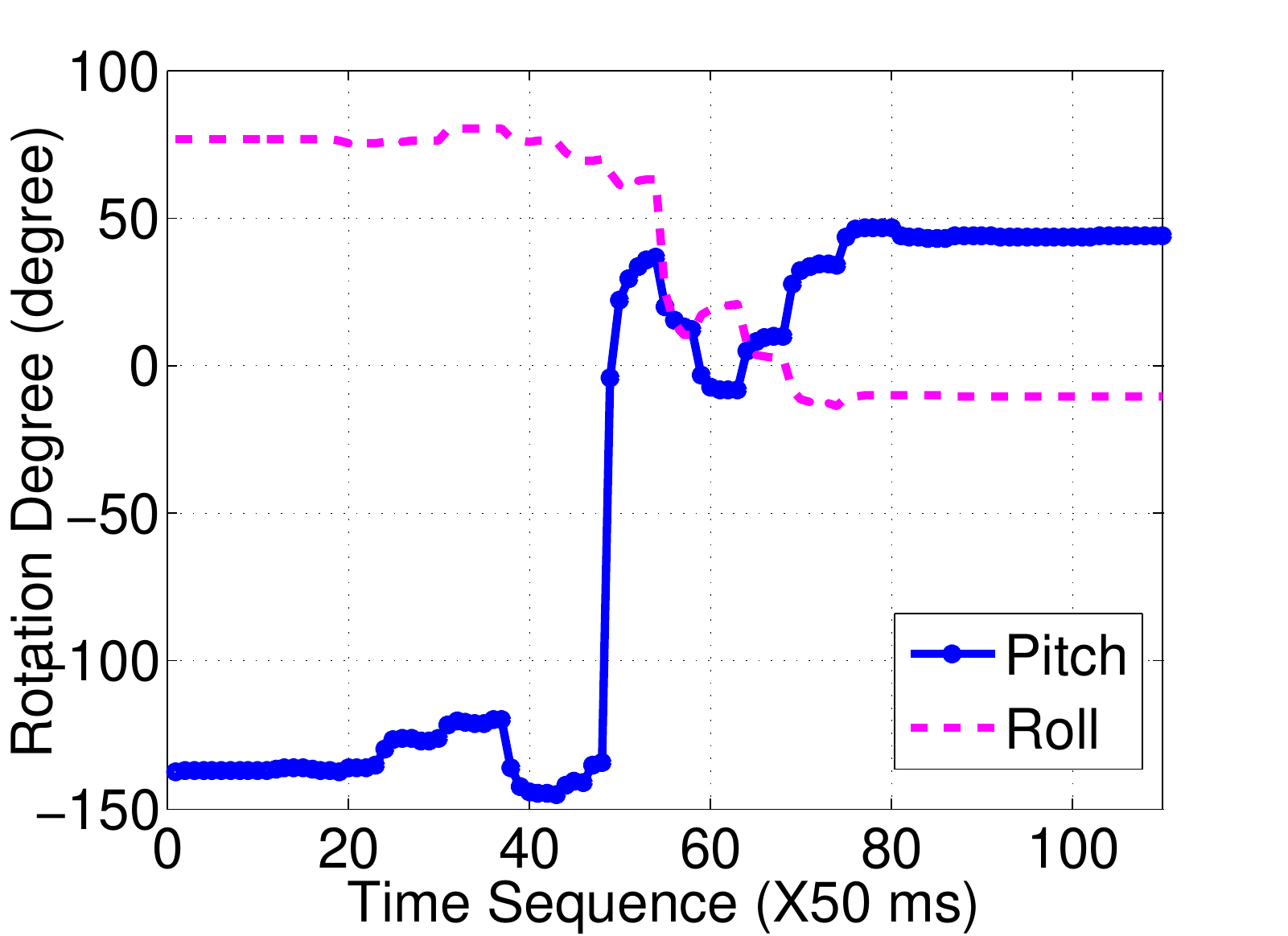}}
%\subfigure[Left Pocket, Passenger Side\label{fig:rl_enter}]{\includegraphics[scale = 0.25]{figures/rl_enter.pdf}}
%\subfigure[Right Pocket, Passenger Side\label{fig:rr_enter}]{\includegraphics[scale = 0.25]{figures/rr_enter.pdf}}
\vspace{-0.1in}
\caption{Side detection: the observation of rotation along BFC.}
\label{fig:rotate}
\vspace{-0.1in}
\end{figure}
\begin{figure*}[!ht]
\centering
\subfigure[Dashboard\label{fig:mag_dashboard}]{\includegraphics[height = 1.2in, width=1.8in]{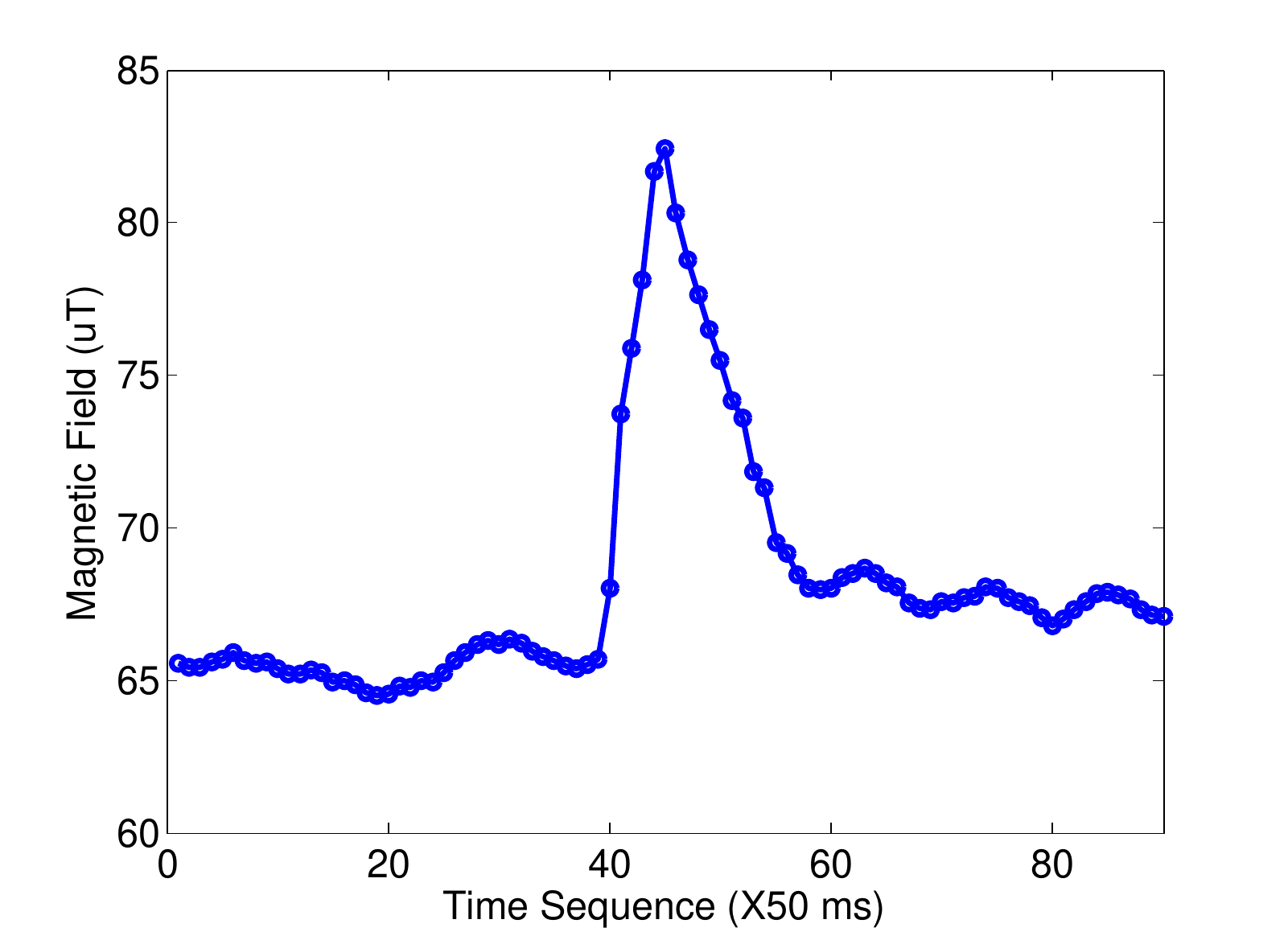}}
\subfigure[Front vs. Back\label{fig:mag_f_b}]{\includegraphics[height = 1.2in, width=1.8in]{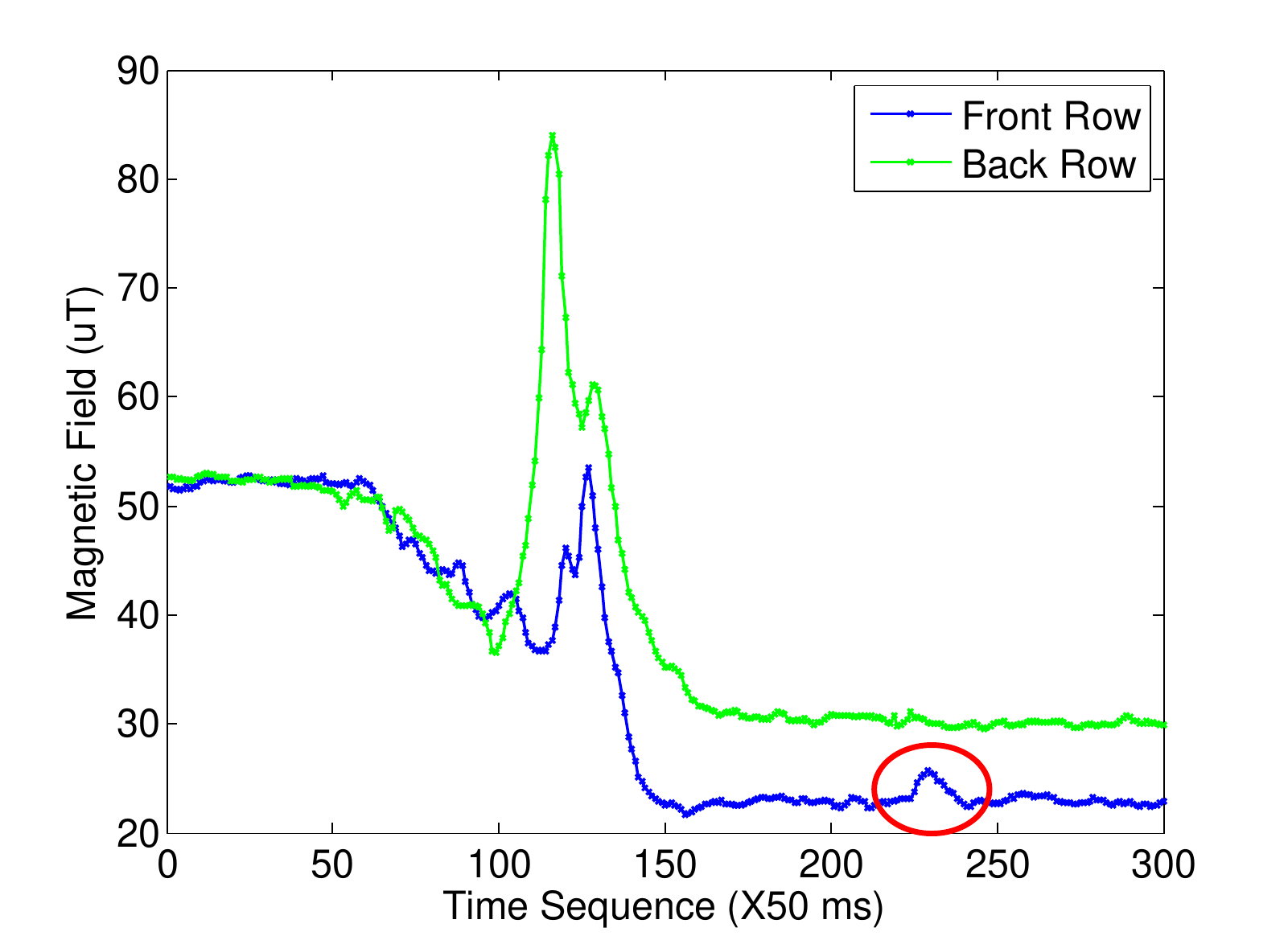}}
\subfigure[Engine Starts\label{fig:mag_f_start}]{\includegraphics[height = 1.2in, width=1.8in]{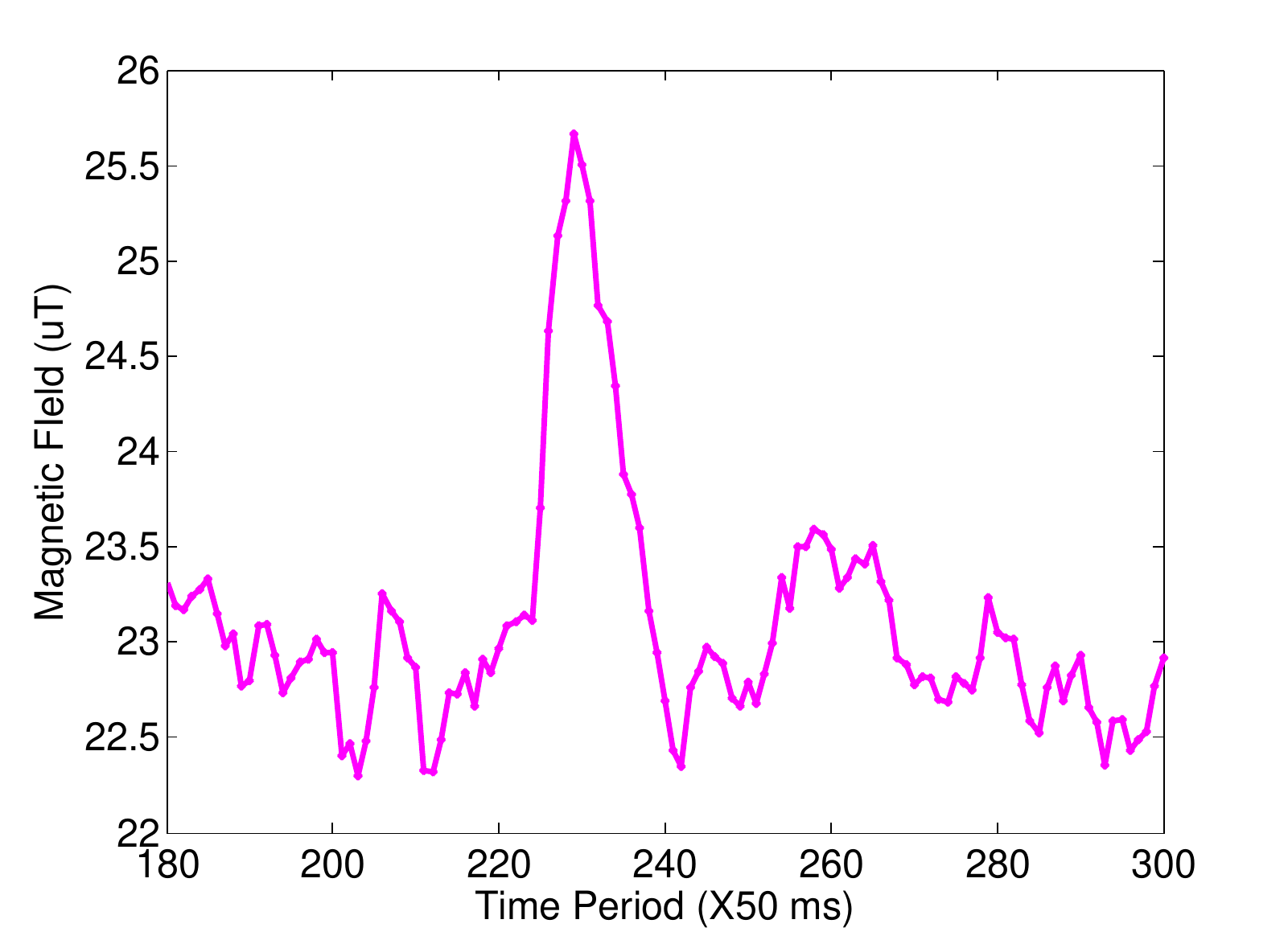}}
\vspace{-0.15in}
\caption{Changes of magnetic field in various scenarios.}
\label{fig:mag_in_car}
\vspace{-0.2in}
\end{figure*}

%Suppose smartphone is located in the trouser pocket, according to our
%  experiment, the orientation of smartphone while getting in vehicle
% from driver side and passenger side are different.
Suppose smartphone is located in the trouser pocket, the orientations
are different if we get in vehicle from different sides.
We calculate the continuous orientation of smartphone while user
 entering the vehicle from both driver and passenger sides
 respectively, and plot the varying of pitch and roll in
 Figure~\ref{fig:rotate}.
In addition, the orientation is modified according to Extend Kalman
 Filter, because of the internal mechanism noise of inertial sensors
 and the measurement noise.
Since the orientation of the vehicle is unknown and unpredictable, we
 only consider the rotation in \emph{Pitch} and \emph{Roll}, which will
 be affected when stepping into the vehicle.
%We denote the leg which is close to the vehicle and with smartphone in
% that pocket as the inner leg.
From Figure~\ref{fig:rotate}, the vibration difference is obvious,
 especially when the inner leg is stepping into the vehicle, so that
 the rotation patterns are different, and the side
 detection is feasible.

\subsection{Front or Back Seats?}
%%%% this subsection will present our methods for detecting whether a
%%%% user sits in the front seats or back seats. Here we assume that
%%%% there are multiple rows. We need to discuss the cases when there
%%%% is only one row or other cases?
%\begin{figure*}[htbp]
%\centering
%\subfigure[Magnetic Field around the Dashboard\label{fig:mag_dashboard}]{\includegraphics[scale = 0.33]{figures/mag_windshield.pdf}}
%\subfigure[Magnetic Field in the Front vs. Back\label{fig:mag_f_b}]{\includegraphics[scale = 0.33]{figures/mag_f_b.pdf}}
%\subfigure[Engine Starts\label{fig:mag_f_start}]{\includegraphics[scale = 0.33]{figures/mag_f_start.pdf}}
%\vspace{-0.1in}
%\caption{Changes of magnetic field in various scenarios.}
%\label{fig:mag_in_car}
%\vspace{-0.2in}
%\end{figure*}

%The third phase of our system is  to  locate the position of the
% smartphone in a vehicle: whether it is in the front  or in the back of vehicle.
The third phase of our system is to solve the front-or-back problem, and by combining
 with the left-or-right result from the second phase, we can locate the exact position
 of the smartphone in a vihicle.
%Actually, available information the smartphone could exploit is limit,
% and it is not easy to identify the location of smartphone inside the
% vehicle.
Actually, useful information that the smartphone can exploit is limited,
 so it is hard to identify the accurate front-or-back location of a smartphone
 inside a vehicle.
The latest work is based on calculating the distances between
 smartphone and the speakers through acoustic
 ranging~\cite{yang2011detecting}, and the other relies on the sound
 level of the turning signal~\cite{chu2011vehicle}.
However, the first solution has to handle the issues of  the
  placement of speakers, and
  the latter needs collaboration between phones and cloud server to do
  the comparison.
In this section, we will introduce two independent approaches (based
 on changes of magnetic field when engine starts, and changes of
 accelerometer when  vehicle passing through bumps and potholes) to
 determine whether the smartphone is located in the front row or the
 back row.

Smartphone is capable of sensing the  magnetic field, and the
 special mechanical structure of the vehicle will affect the
 surrounding magnetic field.
We take a set of further experiments to test the altering condition of
 magnetic field from walking towards the car to  being ready to
 drive in two different cases, one is sitting on the front row, and the
 other is sitting on the back row.
%The tests are taken under the circumstance that smartphone is
% in the trouser pocket, we calculate the joint vector value of magnetic
% field for both cases, and plot the results in Figure~\ref{fig:mag_in_car}.
%\begin{figure*}[!htb]
%\centering
%\subfigure[Magnetic Field around the Dashboard\label{fig:mag_dashboard}]{\includegraphics[scale = 0.33]{figures/mag_windshield.pdf}}
%\subfigure[Magnetic Field in the Front vs. Back\label{fig:mag_f_b}]{\includegraphics[scale = 0.33]{figures/mag_f_b.pdf}}
%\subfigure[Engine Starts\label{fig:mag_f_start}]{\includegraphics[scale = 0.33]{figures/mag_f_start.pdf}}
%\vspace{-0.1in}
%\caption{Changes of magnetic field in various scenarios.}
%\label{fig:mag_in_car}
%\vspace{-0.2in}
%\end{figure*}

We first put the smartphone on the dashboard, where the place is much
 closer to the engine, the value of magnetic field is relatively large,
 around $65uT$ when the engine is off.
After we start engine, as shown in Figure~\ref{fig:mag_dashboard}, the
 value experiences a slight increase to approximately $67uT$ after an
 obvious spike, which reflects a large fluctuation of the magnetic
 field at the very moment of engine starts.
The sudden spike provides us a good signal to detect whether the engine
 starts, with the amplitude nearly increased by $20uT$.
However, most of users may put their smartphones in the cup
 holder or leave them in the pocket, and that signal may not be so obvious
 because of the increasing distance between the smartphone and the
 engine.
Thus we take another test for both sitting in the front row and the
 back row to evaluate the difference of the spike.
The tests are taken in a continuous period: walk towards the
 car, open the door, sit down, and start engine.
 We  plot the
value of the magnetic field  in Figure~\ref{fig:mag_f_b}.
Based on this figure, there are two observations: one is that the
level of magnetic field is similar when the user is away from the
vehicle, and the other is  that the magnetic field in the back
row is \emph{larger} than the front row (which is somewhat counter
intuitive).
An exciting phenomenon is that even the smartphone is in the cup
holder or the trouser pocket, the magnetometer  could still sense the
variation (red circle) of the magnetic field while the engine starting,
but with a smaller amplitude change (around $3uT$, and the zooming in
figure is shown in Figure~\ref{fig:mag_f_start}).
%while sitting in the back row could record nothing.
And we also found that if the smartphone is located on the back seat,
 it will record nothing.
Thus, we exploit the instantaneous
 magnetic field vibration when the engine starts to
 determine the rows by fusing  the readings from accelerometer.
%When the GPS indicates the vehicle is
%  moving, if we detect the vibration we could ensure that with high
% probability we are sitting in the front row, and if the magnetic field
% is  stable  then with high possibility we are sitting in the back
% row.
When the vehicle is moving has been indicated, our system will look through
 the stored data in the buffer,
 and if the magnetic vibration can be detected, we know that the smartphone is located in the front
 row with high probability.

Since the system is running according to specific duty cycle, and
chances that the mis-detection is possible.
Therefore, the second method is now proposed.
In both Nericell~\cite{mohan2008nericell} and Pothole Patrol
(P$^{2}$)~\cite{eriksson2008pothole}, researchers use acceleration
sensors and GPS deployed on vehicles to detect the pothole and bump on
the road.
And the pothole and bump will result in significant vertical spikes
and dips in acceleration in the gravity direction, and machine
learning approach is adopted to identify these.

Empirically, when we drive through a bump or a
pothole, people sitting in the back row feel more bumpy than those sitting in
the front row.
%In order to prove this, we collect a set of data by driving through
%both bumps and potholes to match the sensory data to the real feeling.
We try to use this phenomenon as a smart evidence for detecting front-or-back.
We collect a set of data by driving through both bumps and potholes to match the
sensory data to the real feeling.

%\begin{figure}[!htb]
%\centering
%\subfigure[Driving through bump\label{fig:bump_f_b}]{\includegraphics[scale = 0.25]{figures/bump_f_b.pdf}}
%\subfigure[Driving through pothole\label{fig:pothole_f_b}]{\includegraphics[scale = 0.25]{figures/pothole_f_b.pdf}}
%\vspace{-0.1in}
%\caption{Driving through bumps and potholes.}
%\label{fig:bump_pothole}
%\vspace{-0.2in}
%\end{figure}
%\begin{figure*}[htb]
%\centering
%\subfigure[Time Interval between two inputs\label{fig:cdf_texting}]{\includegraphics[scale = 0.25]{figures/cdf_texting.pdf}}
%\subfigure[Frequency of typing\label{fig:info_texting}]{\includegraphics[scale = 0.25]{figures/info_texting.pdf}}
%\subfigure[The probability of different time interval between two inputs\label{fig:pdf_texting}]{\includegraphics[scale = 0.25]{figures/pdf_texting.pdf}}
%\subfigure[The typo frequency\label{fig:cdf_typo}]{\includegraphics[scale = 0.25]{figures/cdf_typo.pdf}}
%\vspace{-0.15in}
%\caption{The information extracted from typing.}
%\label{fig:typing_norm}
%\vspace{-0.2in}
%\end{figure*}
\begin{figure}[!htb]
\vspace{-0.1in}
\centering
\subfigure[Driving through bump\label{fig:bump_f_b}]{\includegraphics[scale = 0.25]{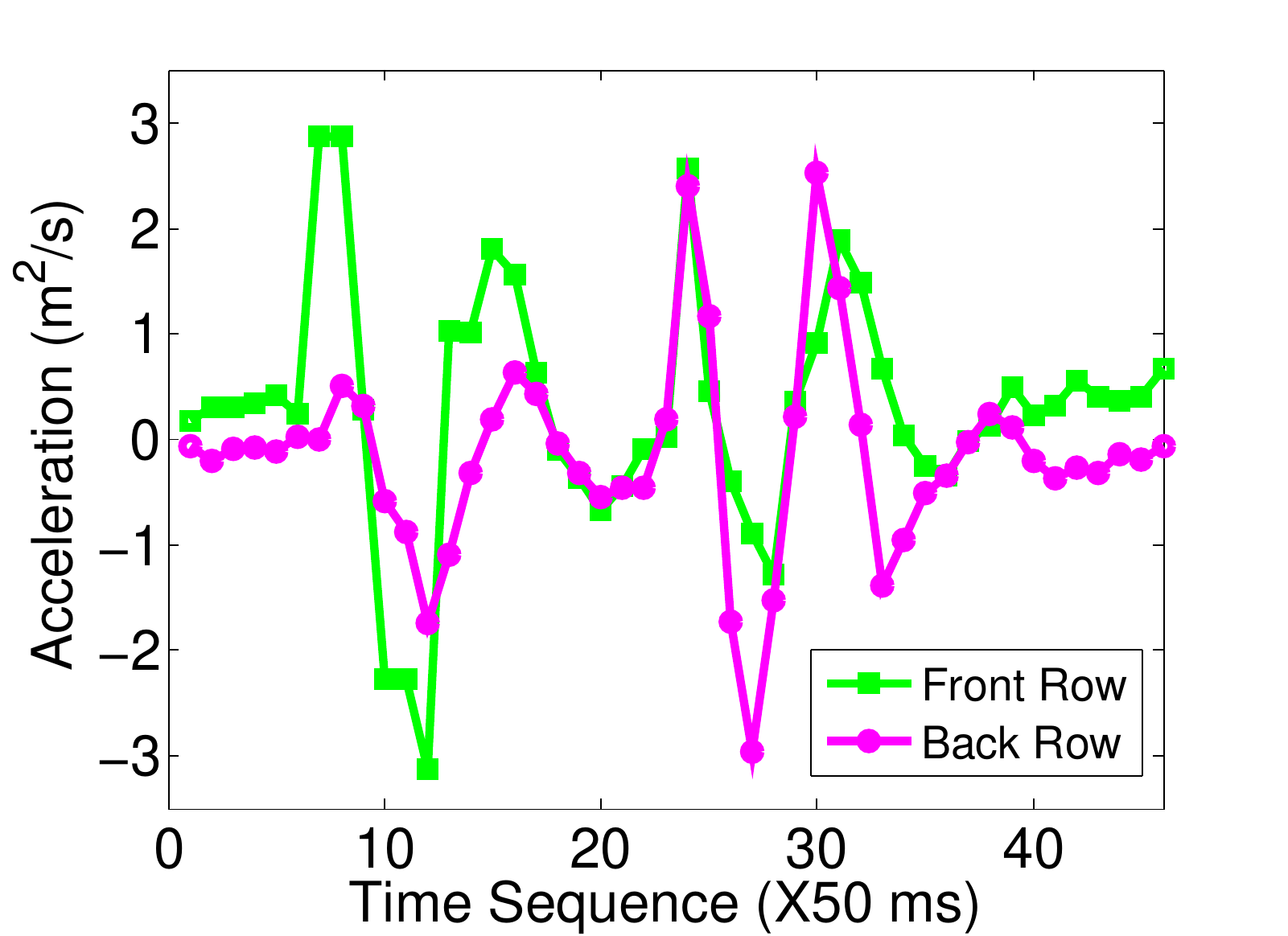}}
\subfigure[Driving through pothole\label{fig:pothole_f_b}]{\includegraphics[scale = 0.25]{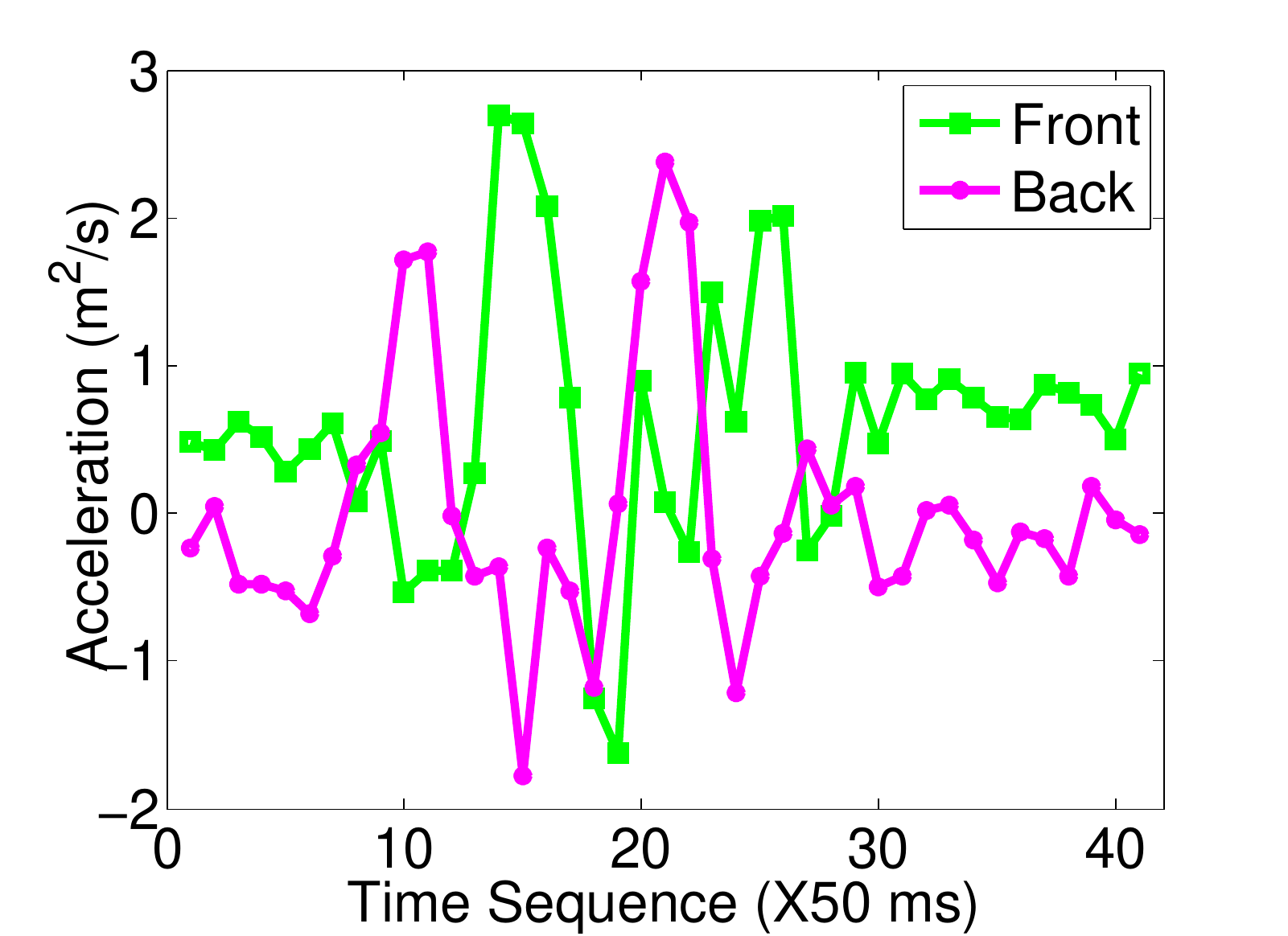}}
\vspace{-0.1in}
\caption{Driving through bumps and potholes.}
\label{fig:bump_pothole}
\vspace{-0.1in}
\end{figure}
We deploy two smartphones in two passengers, sitting on the front and
 back row respectively, and sample the acceleration in $20Hz$ during the driving.
The accelerations are converted into the gross linear acceleration in
 the horizontal plane, and the ground direction in EFC is in
 Figure~\ref{fig:bump_pothole}.
Generally, the smartphone will observe the road condition twice
 because both front and back wheels will drive through the bump/pothole,
 and the amplitudes are completely different.
Due to the special shape of bump or deceleration strip, one wheel will
 experience two continuously large vibrations, one is first hitting the
  bump and jumping to the highest, and the other is hitting the ground
 after driving through.
\begin{figure*}[htb]
\centering
\subfigure[Time Interval between two inputs\label{fig:cdf_texting}]{\includegraphics[scale = 0.25]{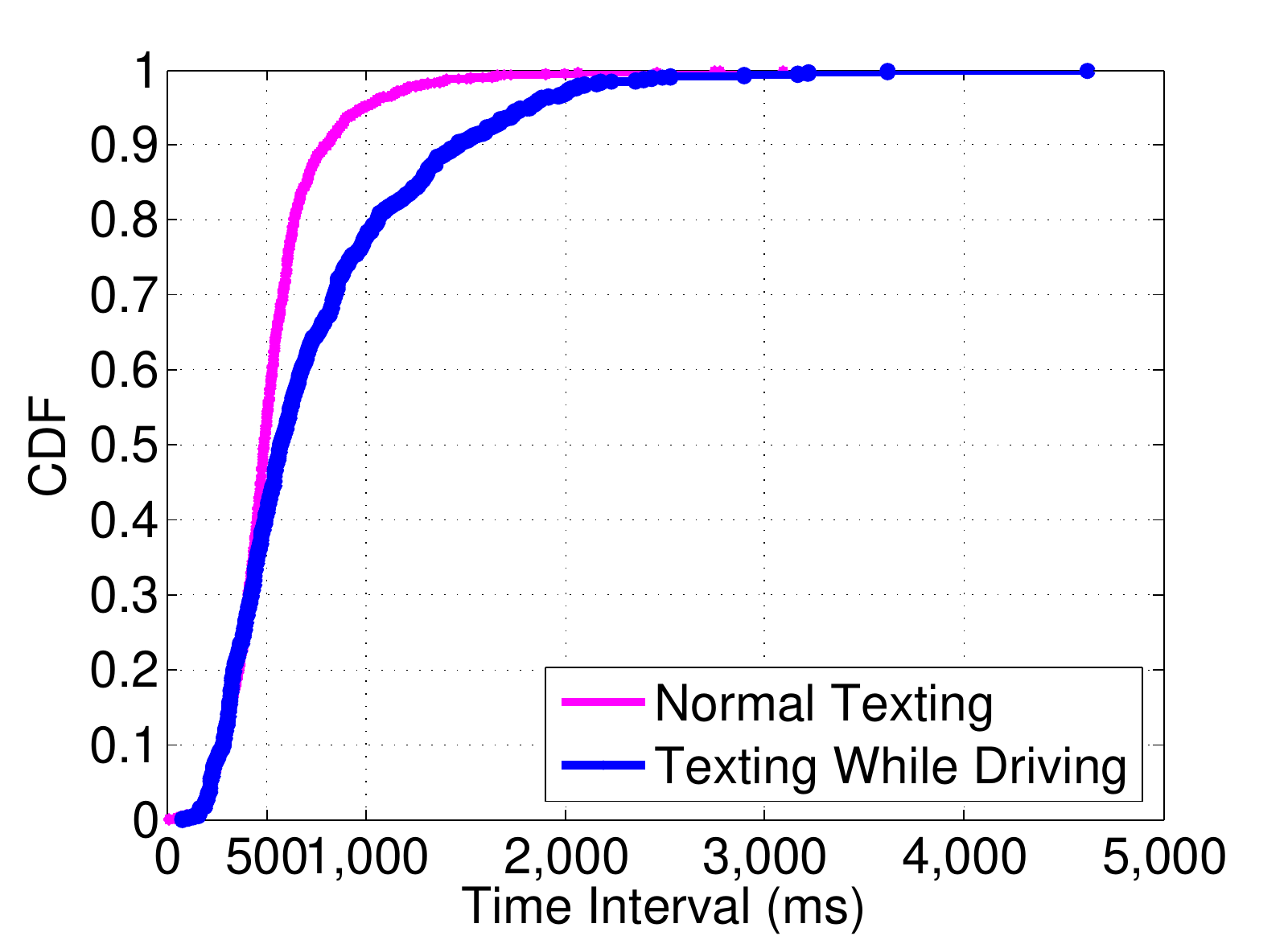}}
\subfigure[Frequency of typing\label{fig:info_texting}]{\includegraphics[scale = 0.25]{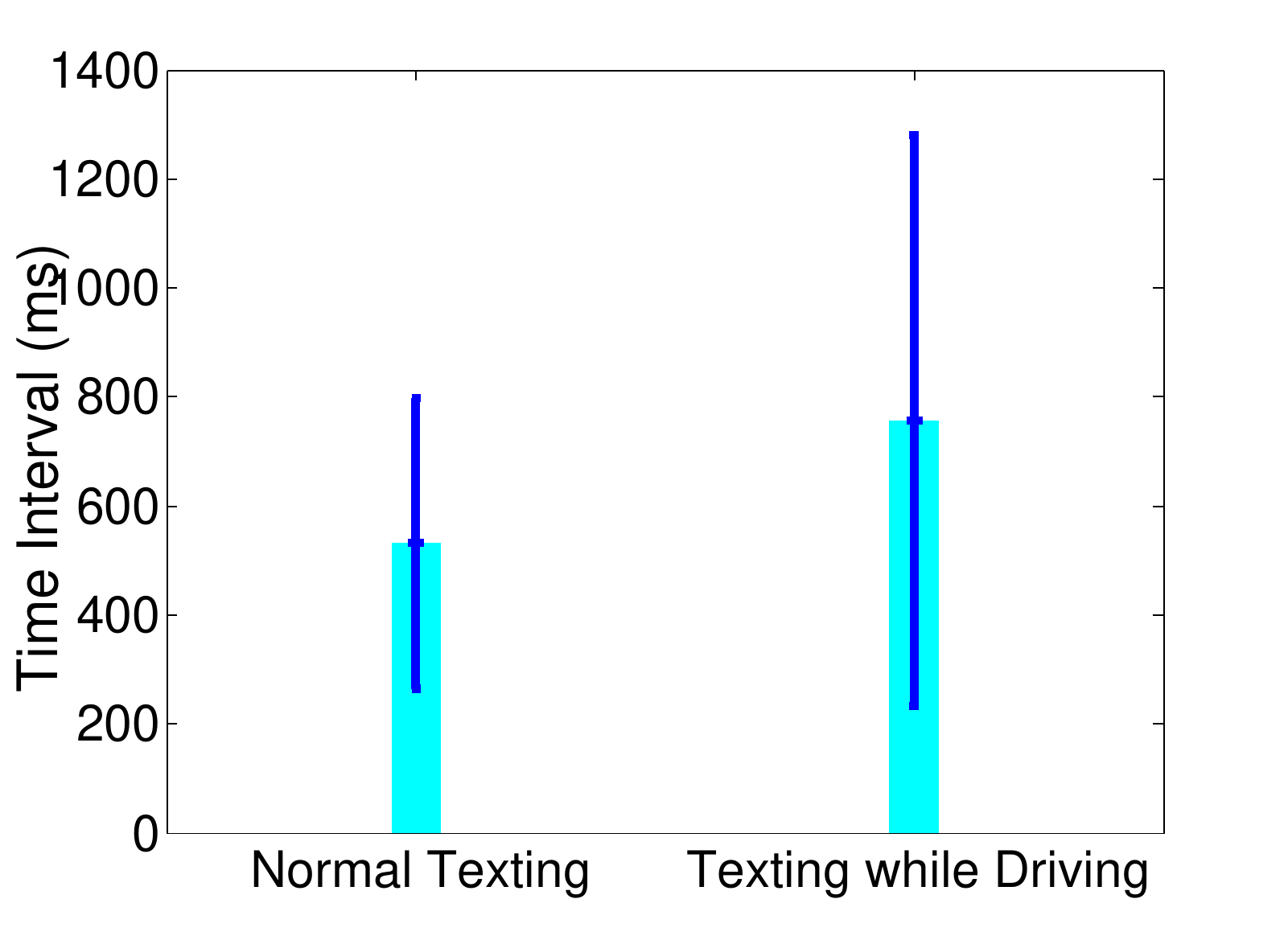}}
\subfigure[The probability of different time interval between two inputs\label{fig:pdf_texting}]{\includegraphics[scale = 0.25]{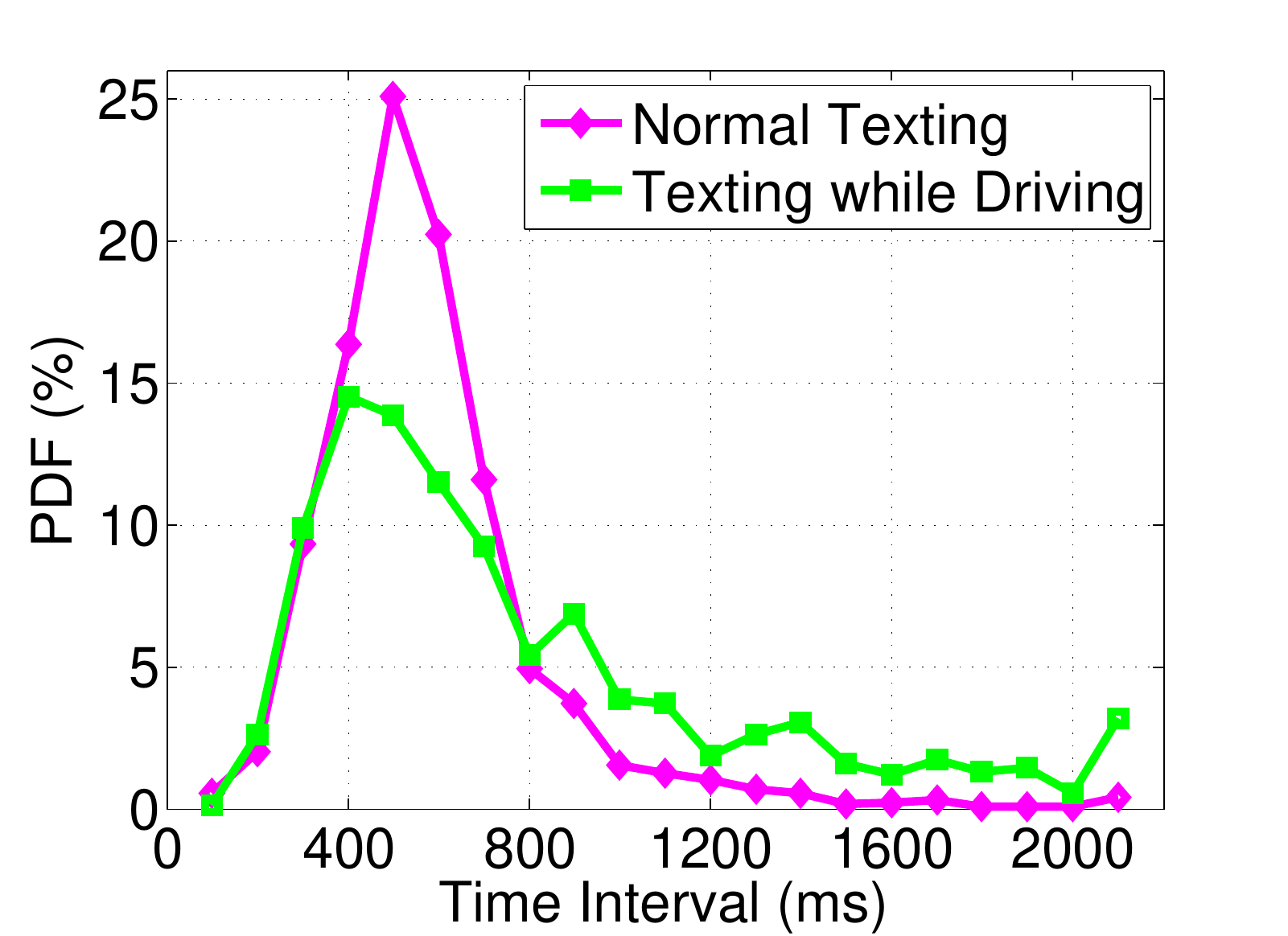}}
\subfigure[The typo frequency\label{fig:cdf_typo}]{\includegraphics[scale = 0.25]{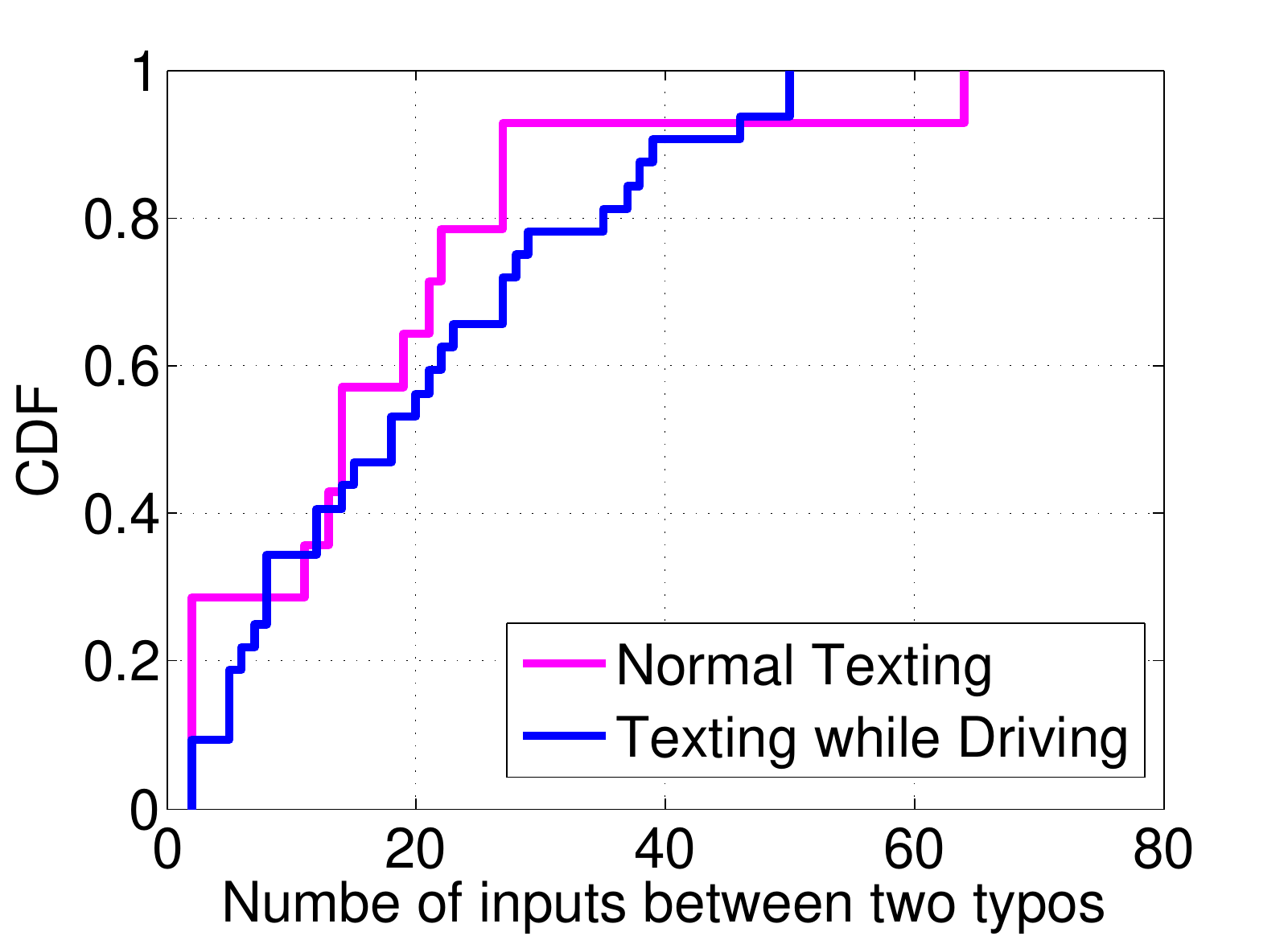}}
\vspace{-0.15in}
\caption{The information extracted from typing.}
\label{fig:typing_norm}
\vspace{-0.2in}
\end{figure*}
When the smartphone is in the front row, the intensity
 difference between two continuous bumps is relatively small, while in
 the other case, the difference is much larger.
The smartphone could only detect a small jump when the front wheel
passes the bump, but the back wheel will make the jump much higher, as
shown in Figure~\ref{fig:bump_f_b}.
The pattern is similar when it comes to the pothole as in Figure~\ref{fig:pothole_f_b}.
In real scenario, most of the cars will vibrate during driving even if
 the road is smooth enough.
When vehicles drive on a bumpy road, both front and back wheel will
 experience a sudden drop and then jump or a jump followed by a drop.

\subsection{Texting?}
%\begin{figure*}[htb]
%\centering
%\subfigure[Time Interval between two inputs\label{fig:cdf_texting}]{\includegraphics[scale = 0.25]{figures/cdf_texting.pdf}}
%\subfigure[Frequency of typing\label{fig:info_texting}]{\includegraphics[scale = 0.25]{figures/info_texting.pdf}}
%\subfigure[The probability of different time interval between two inputs\label{fig:pdf_texting}]{\includegraphics[scale = 0.25]{figures/pdf_texting.pdf}}
%\subfigure[The typo frequency\label{fig:cdf_typo}]{\includegraphics[scale = 0.25]{figures/cdf_typo.pdf}}
%\vspace{-0.15in}
%\caption{The information extracted from typing.}
%\label{fig:typing_norm}
%\vspace{-0.2in}
%\end{figure*}
While driving, the accident will be more likely to happen when the
 driver is distracted, such as texting, twittering and composing
 email.
In order to prevent the driving distraction, the second function of
 the system is to detect texting during driving.
Once the endangered behavior is detected, the system could alert the
 driver or the relatives through network~\cite{rode}.

Generally, typing is not a difficult task when user are fully
 concentrated, with fewer typo and higher accuracy.
While user typing on the smartphone but partly distracted, the time
 interval between words or letters may last much longer, and the typing
 accuracy may be much lower, which leads to the two criteria to determine
 if the user is typing in a normal manner or distracted manner.
The criteria of determining whether the user is fully concentrated  or
distracted from texting depends mainly on the frequency of typing and
the probability of typo appears.
We conduct a set of experiments by a group of colleagues to compose
multiple sentences in the smartphone in both driving and normal
condition.
Due to the safety issue, the driving scenario is conducted in the parking lot.
In this initial experiment, we record both the time interval between
consecutive inputs of letter, and the number of letters between two
consecutive typos.

We plot the CDF of both cases in Figure~\ref{fig:cdf_texting}, and the
general statistic information in Figure~\ref{fig:info_texting}.
In the normal texting cases, the user usually is fully concentrated so
that the typing speed is relatively higher than that of the abnormal
cases.
Thus about $90\%$ of typing inputs falls within $800ms$ in the former
scenario, while the same interval only covers less than $70\%$ inputs
in the latter scenario.
Based on statistic information, the average time interval is around
$536.55ms$ with Standard Deviation $327.03ms$ for the normal scenario
and the value in the distracted scenario is $742.42ms$ and $528.68ms$
respectively.
Typing while driving, people usually type one word or phrase, and
 then pause for a while to watch the road before continue typing,
 such special behavior leads to the large standard deviation of input
 interval in the distracted scenario.
Such behavior habit also results in the fact that there is still a
 certain proportion of inputs with interval less than $500ms$, as shown
 in the Probability Density Function (PDF)  in
 Figure~\ref{fig:pdf_texting}.
Simultaneously, the amount of typos in the distracted scenario is much
 larger than the normal scenario.
We compute the amount of inputs between two continuous typos,
 generating from backspace and the CDF is shown in
 Figure~\ref{fig:cdf_typo}.
Generally, typo appears in approximately every $50$ inputs in normal
 condition, while only $30$ inputs in the distracted scenario.

\section{Reducing Energy Consumption}

%The basic strategy adopted by \ourprotocol to reduce energy
% consumption is dynamic sampling according to self-learnt daily routine
% generated through a close loop, as shown in Figure~\ref{fig:process}.
The basic strategy adopted by \ourprotocol to reduce energy
 consumption is dynamic sampling according to self-learnt daily routine
 generated through a close loop.
%\begin{figure}[hptb]
%\centering
%\includegraphics[scale=0.4]{figures/strategy.pdf}
%%\includegraphics[scale=0.4]{figures/system_loop2.pdf}
%\vspace{-0.1in}
%\caption{The System Loop}
%\label{fig:strategy}
%\end{figure}
Based on our observation, we notice that users often
 have to walk to the parking lot or the garage before getting on the
 vehicle.
In addition, most of the users drive at some fixed time everyday.
Thus in our strategy, \ourprotocol starts with walking
 %detection according to two separate timelines: commute-time (\eg, at 8AM
 %in morning and 5PM in afternoon), and duty cycle (\eg, sample every 5
 %minutes).
 detection with different duty cycle (\eg, sample every 5
 minutes) according to the commute-time (\eg, at 8AM
 in morning and 5PM in afternoon).
In commute-time, the system will sample historical data and learn when
 a user will drive in a day.
When the current time $T$ is close to commute-time $T_D$, \ie,
 $T=T_D- \alpha \cdot T_{th}$ where $T_{th}$ is the variance of historical
 commute-times and $\alpha$ is a constant,
 we will sample data
 with large frequency, say $1/t_c$.
In the rest of the time, the system will detect the walking
 activity with sample frequency $1/t_d$ times, which is much smaller
 than $1/t_c$.

\begin{figure}[hptb]
\centering
\includegraphics[scale=0.4]{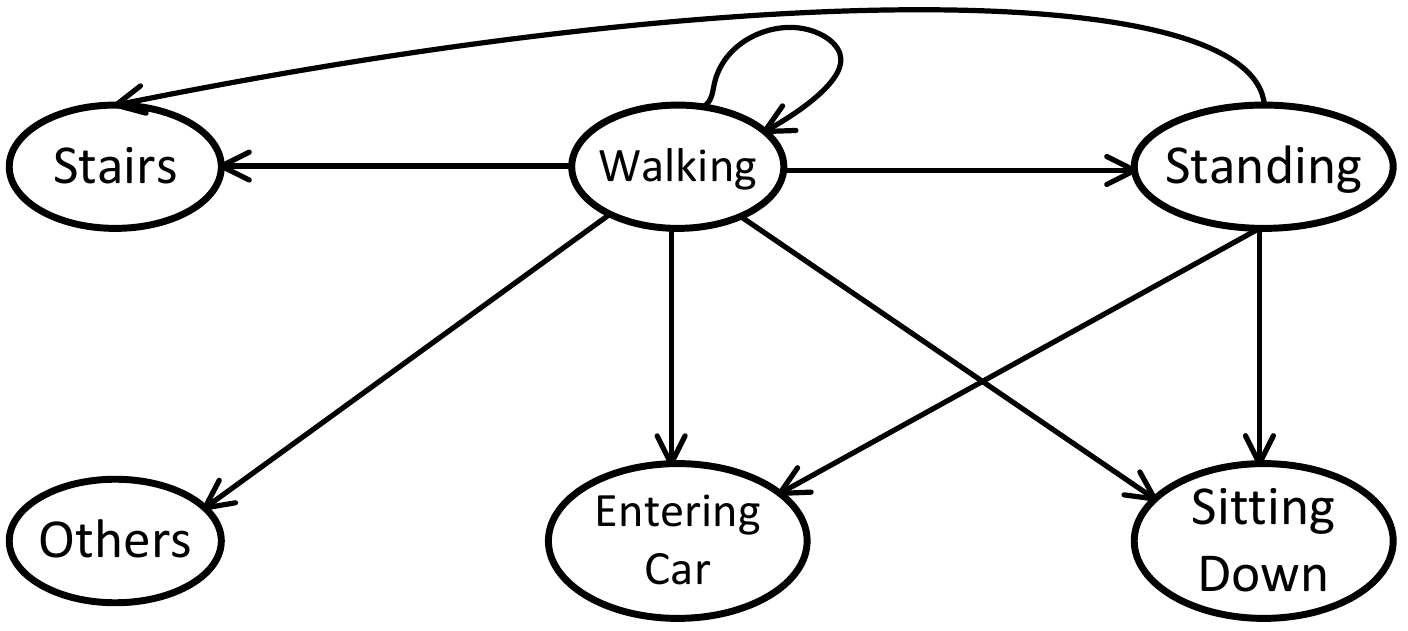}
\vspace{-0.2in}
\caption{The Action Loop}
\label{fig:hmm}
\vspace{-0.1in}
\end{figure}

\ourprotocol learns the habit of user, and adjusts the detecting periods
 automatically based on possibility that a user may start driving.
We record the time of each of the behaviors for certain users for a week,
   including walking, standing, sitting down, driving, ascending
   stairs, and others.
We then model the transition probability between different activities
using Hidden Markov Model (HMM)~\cite{rabiner1986introduction} as
shown in Figure~\ref{fig:hmm}.
From  statistics of a week data, we calculate the initial probability
 of each activity as shown in Table~\ref{table:init_act}.

\begin{table}[hptb]
%\vspace{-0.2in}
\centering
\caption{Probabilities of different activities.}
\begin{tabular}{|l|l|l|l|}
\hline
  Behavior & Probability&
  Behavior &  Probability\\
\hline
  Walking & 42.25\% &
  Enter Car & 14.08\%\\
  Stand & 9.86\%    &
  Sit Down & 11.27\%    \\
  Stairs & 15.49\%  &
  Others & 7.05\%   \\
\hline
\end{tabular}
%\vspace{-0.1in}
%\caption{Probabilities of different activities.}
\label{table:init_act}
%\vspace{-0.2in}
\end{table}

We then calculate the transition probabilities from one state to
another state and found that such probability is as high as $16.67\%$
among the behavior we detected.
We put more emphasis on the sampling strategy during the possible
routine driving period, including walking towards the car.
Suppose the time duration before entering the car is $T$, and it could
 be divided into small detecting time slot, denoted as $t_i$, and the
 sampling frequency is $f_i$.
Our goal of deciding the sampling strategy is to minimize the
 overall energy consumption while guarantee the expected
 behavior miss ratio is less than a threshold $\varepsilon$.
In \ourprotocol, we use the following sampling strategy.
Assume that the mean time of walking towards the car is $T$ and
variance is $\sigma$, and we
 have detected walking activity using HMM.
Then we start looking for entering car activity by sampling data and
 performing detection algorithm with time interval
 $t_i = (T-\sigma)\cdot{(\frac{1}{2}})^{i}$, for $i=1, 2, \cdots$.

We then study the energy consumption if we need to use bump
and/or pothole signal for driver detection.
Suppose the vehicle is driving at a constant velocity, and the bumpy
 detection is taken in a cycle $w + s$, where $w$ is the duration of
 detection and $s$ denotes the sleep.
\begin{figure}[!ht]
\vspace{-0.1in}
\centering
\includegraphics[height=1.5in,width=2in]{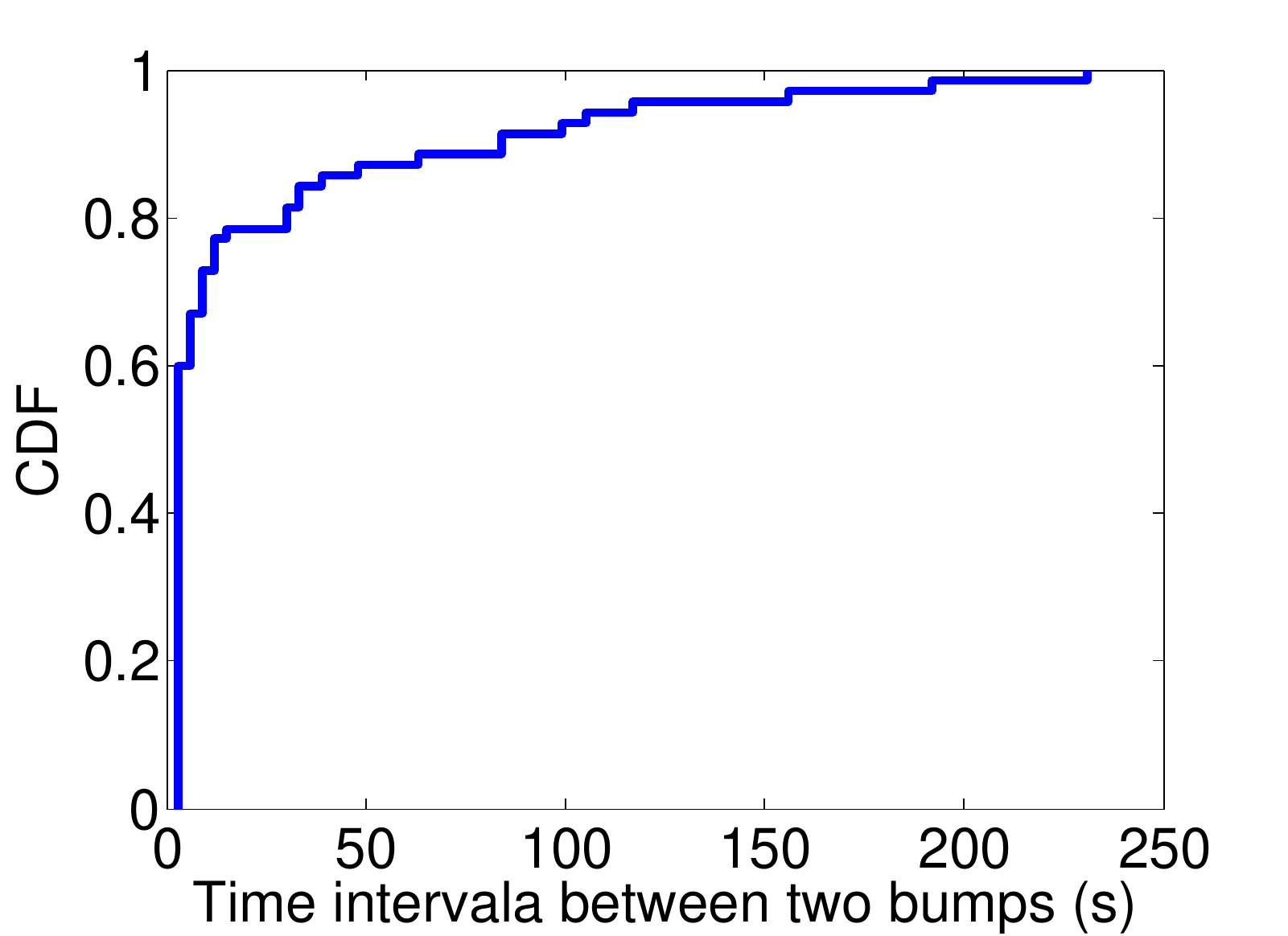}
\vspace{-0.15in}
\caption{Bump on the road}
\label{fig:bump_road}
\vspace{-0.15in}
\end{figure}
%In Figure~\ref{fig:bump_hit}, the blue chunk indicates the system
%collecting data and checking the bump or pothole, while the grey patch
%denotes the bump and pothole.
The system will stop checking until the system detects the
existence of the  bump or pothole.
If bumps and potholes follow  Poisson Process, the probability of
 detecting $k$ bump or pothole in time interval $[t, t + \tau]$ is:
$P(k) = \frac{e^{-\lambda\tau}{\lambda\tau}^{k}}{k!}$
where $\lambda$ is a rate parameter.
Thus, the probability of successfully detecting the $k$th bump or pothole
 by the $i$th detecting cycle is:
$P_{ik} = P_{ith~hit} \cdot {P_{k-1~miss}}
        = (1 - e^{-w\lambda})\cdot{\frac{s}{s + w}}$.
Suppose the average power for sampling sensory data and running
 activity recognition in one unit time is $C$, as a result, the total
 energy consumption under the same circumstance is $C((i - 1)(w + s) +
 t)$, where $t$ is the time for identifying a bump or pothole in the
 $i$th sampling.
And the overall expected cost is
$E(k) = (1 - e^{-w\lambda})\cdot{\frac{s}{s + w}} \cdot {C((i - 1)(w +
  s) + t)}$.
We  test a segment of the road (over $5$ miles), containing both local streets and
highway.
The actual "bump" measured in our data is not the regular speed bump people experience.
We treat any non-smoothy part of a road segment that will cause "bump-like" behavior as a bump,
 and record the time interval of driving through a bump or
pothole on the street as shown in Figure~\ref{fig:bump_road}.
%\mynote{need to show how this data is used}
The figure shows that the probability of a vehicle driving through a
 bump within $50$ seconds is over $80\%$, so that method is feasible and reliable.

%%%%%%%%%%%%%%%%%%%%%%%%%%%%%%%%%%
\section{Evaluations}
In our evaluation, we use both Samsung Galaxy S$3$ (Phone $1$) and
Galaxy Note \uppercase\expandafter{\romannumeral2} (Phone $2$) with
Android $4.1.2$ as the platform.
Since the driver detection consists of three steps, and we will evaluate
 each step separately.
The whole process is evaluated on street in Chicago, except the texting part
 is evaluated in a large parking lot.
To study the impact of different users, we also evaluate
 the system by different users.

\subsection{Getting on Vehicle}
\begin{figure}[!ht]
\centering
\subfigure[Recognition of entering vehicles\label{fig:activity_res}]
{\includegraphics[scale=0.25]{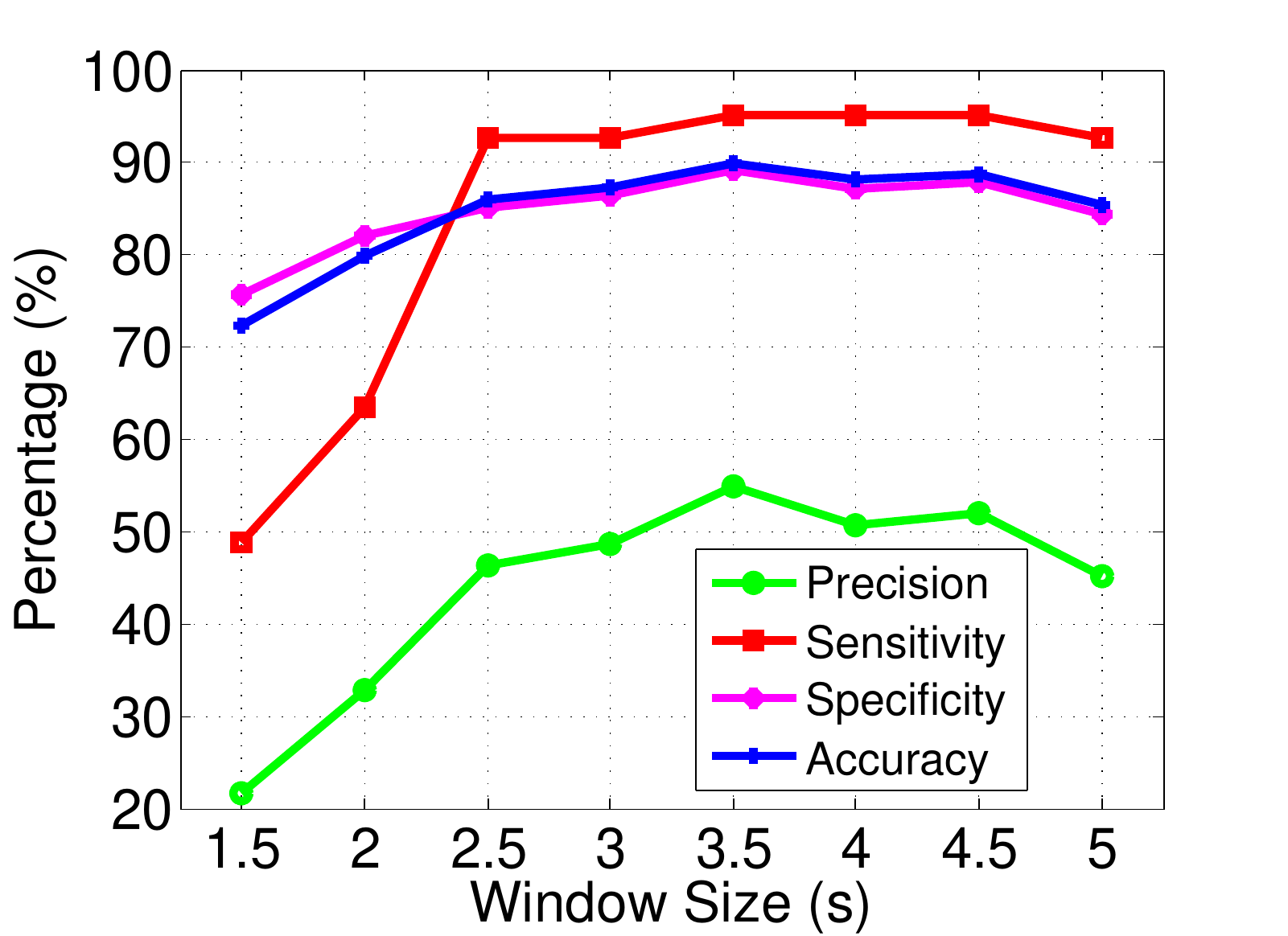}}
%\subfigure[Location of phone in pocket\label{fig:lr_pocket_res}]
%{\includegraphics[scale=0.33]{figures/lr_pocket_res.pdf}}
\subfigure[Detecting the first arriving signal\label{fig:sequence_enter}]
{\includegraphics[scale=0.25]{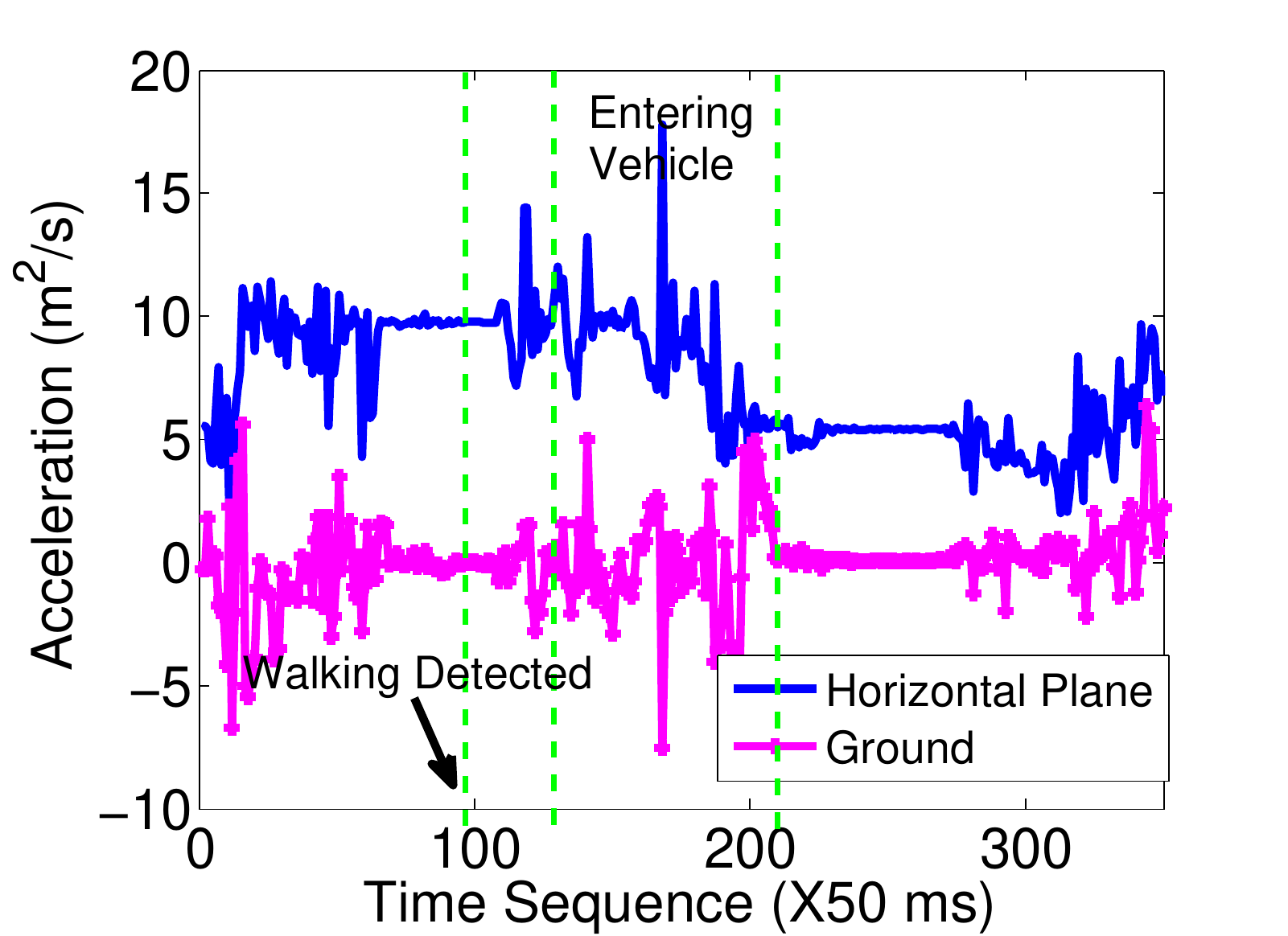}}
\vspace{-0.15in}
\caption{Detecting entering vehicles.}
\vspace{-0.1in}
\end{figure}

Our initial evaluation is the performance of the activity detection,
more specifically, the capability of extracting the behavior of
entering vehicles from large amount of activities.
We run a week-long experiment to gather the information of user's
behavior regularity, notice that most of our colleagues drive only on
commute time.
Since the system is running in the background, it will detect multiple
activities throughout a day besides entering vehicles.
We collected totally $41$ behaviors of entering vehicle in both SUV
 and sedan as well as $296$ other activities, and the result of
 precision, sensitivity, specificity and accuracy are plotted in
 Figure~\ref{fig:activity_res}.
% \begin{figure*}
%  \begin{minipage}[t]{0.4\linewidth}
%    \centering
%    \subfigure[Recognition of entering vehicles\label{fig:activity_res}]
%        {\includegraphics[scale=0.2]{figures/activity_res.pdf}}
%    \subfigure[Detecting the first arriving signal\label{fig:sequence_enter}]
%        {\includegraphics[scale=0.2]{figures/sequence_enter.pdf}}
%    \vspace{-0.1in}
%    \caption{Detecting entering vehicles.}
%    \vspace{-0.2in}
%  \end{minipage}%
%  \begin{minipage}[t]{0.6\linewidth}
%    \centering
%    %\includegraphics[width=1.5in]{graphic.eps}
%    \subfigure[Accuracy in Left vs. Right\label{fig:win_lr}]
%        {\includegraphics[scale=0.2]{figures/win_lr.pdf}}
%    \subfigure[Side Detection\label{fig:side_res}]
%        {\includegraphics[scale=0.2]{figures/side_res.pdf}}
%    \subfigure[New Training Data\label{fig:side_new_train}]
%        {\includegraphics[scale=0.2]{figures/side_new_train.pdf}}
%    \vspace{-0.1in}
%        \caption{Side detection accuracy in different window sizes.}
%        \label{fig:win_side}
%        \vspace{-0.2in}
%  \end{minipage}
%\end{figure*}%
\begin{figure*}[!ht]
\vspace{-0.1in}
\centering
\subfigure[Accuracy in Left vs. Right\label{fig:win_lr}]{\includegraphics[height = 1.2in, width=1.8in]{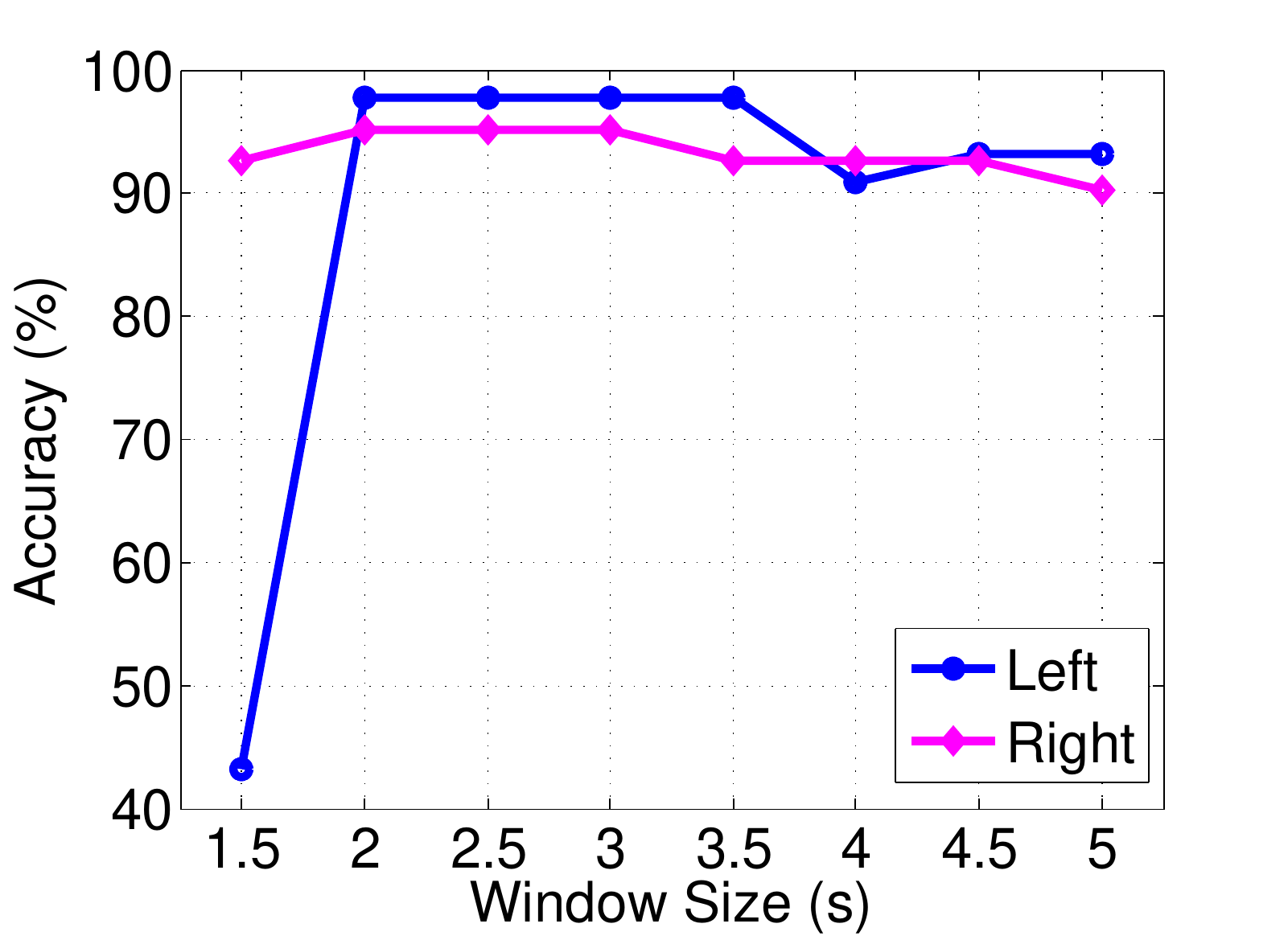}}
\subfigure[Side Detection\label{fig:side_res}]{\includegraphics[height = 1.2in, width=1.8in]{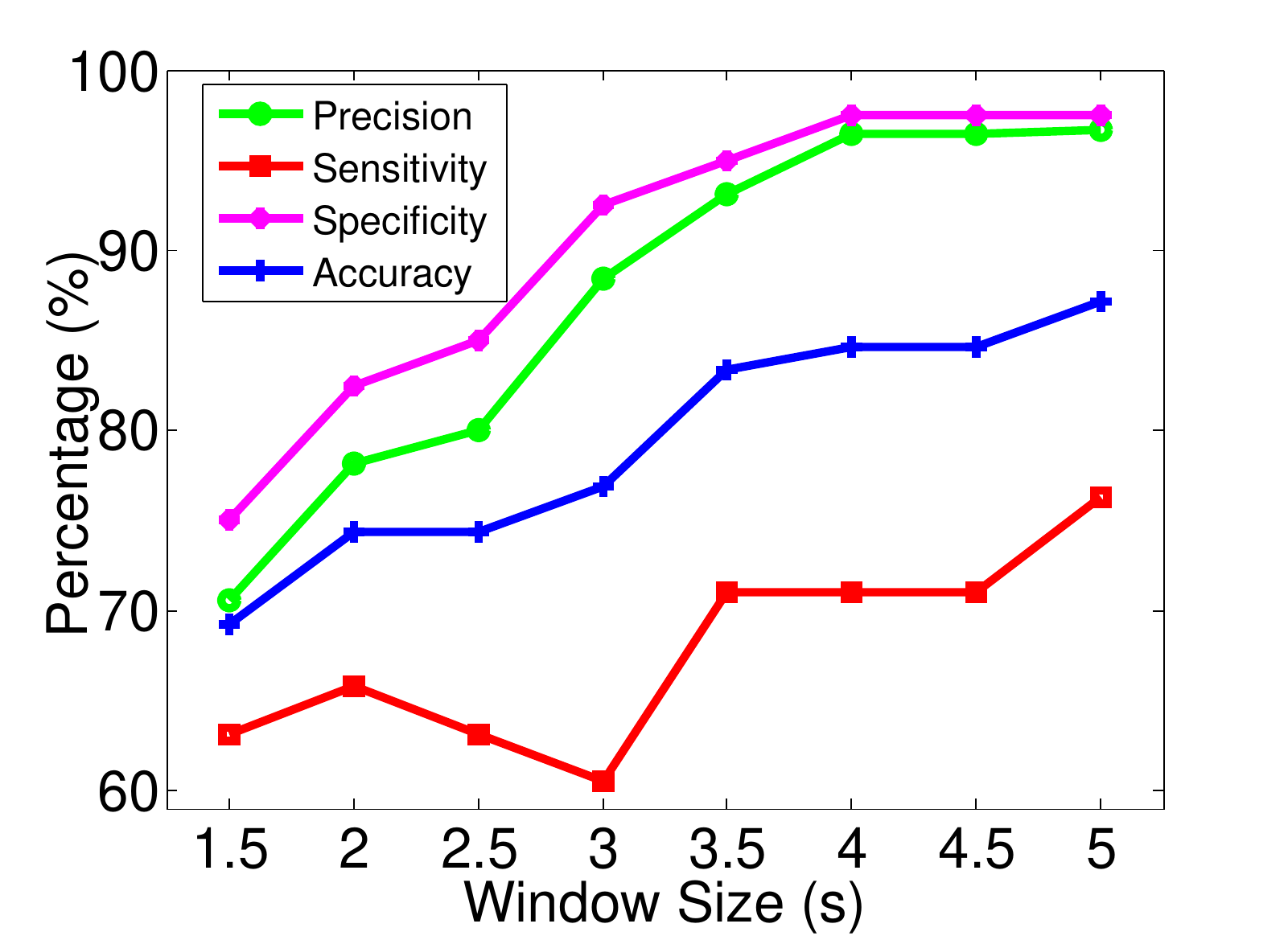}}
\subfigure[New Training Data\label{fig:side_new_train}]{\includegraphics[height = 1.2in, width=1.8in]{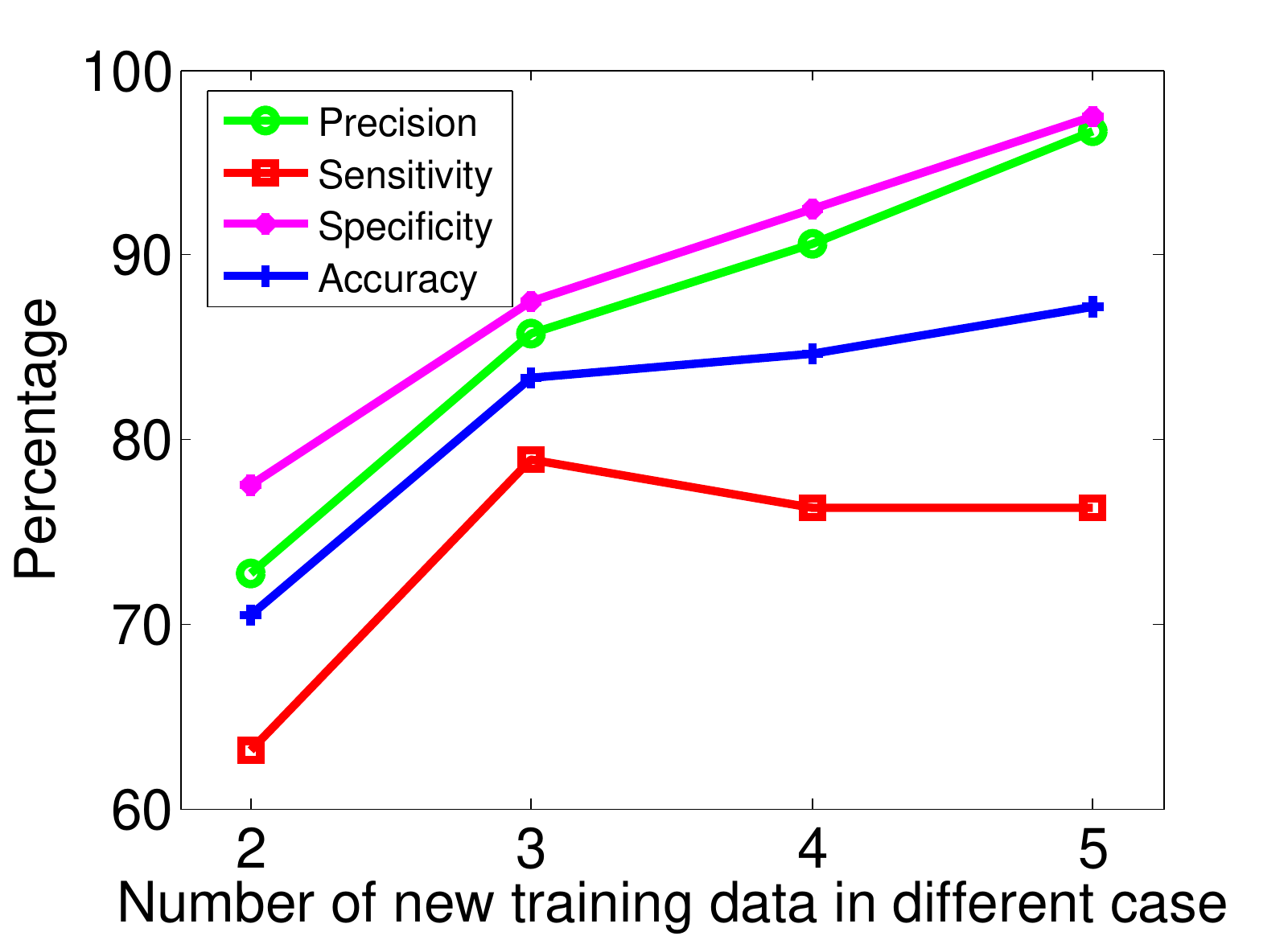}}
\vspace{-0.1in}
\caption{Side detection accuracy in different window sizes.}
\label{fig:win_side}
\vspace{-0.2in}
\end{figure*}
We  test the performance in different window size, ranging from
$1.5s$ to $5s$.
As it shows in the figure, the performance improves with the increase
of window size.
When the window size is around $4s$ to $4.5s$, the results are
 similar, the sensitivity of both cases are over $90\%$.
It is worth mentioning that the precision in all the cases are not as
 high as expected, the reason is because the number of difference between
 true positive and false positive are not large enough.
For example, the true positive and false positive of both cases are
$39 to 38$ and $39 to 36$ respectively.
%Such bottleneck comes from the the fact that other similar behaviors are also
% considered as entering vehicle.
%Therefore, we introduce the changing of acceleration
% as additional feature filter to reduce the false positive.
%If the the behavior is considered as entering vehicle, the smartphone will sense a momentarily
% acceleration fluctuation later, indicating the user is in a moving vehicle.
In addition, the specificity of the cases of window size being larger
 than $3.5s$ are close to $90\%$, and the accuracy is similar.
%However, even if such circumstance happens, the system still could
%determine if the user is in the vehicle or not through other filters,
%such as acceleration when the car is moving, the magnetic field
%varying when entering the car.
%The figure also indicates that, when we choose the window size $4s$ to
%$4.5s$, the accuracy is close to $90\%$.
%However, we discover that the main bottleneck comes from the precision
%because other similar behaviors have been considered as entering vehicle.
%Therefore, we introduce an additional feature of driver determination
%as a filter to reduce the false positive.
After entering vehicle, the smartphone may sense both momentarily magnetic fluctuation
 and acceleration fluctuation, which demonstrates the users  being in a moving car.
According to such idea, we improve our method, and evaluate the performance.
Surprisingly, the value of false positive decreases to zero, thus both the
precision and specofocoty increases to $100\%$.

%\begin{figure*}[!ht]
%\centering
%\subfigure[Recognition of entering vehicles\label{fig:activity_res}]
%{\includegraphics[scale=0.33]{figures/activity_res.pdf}}
%\subfigure[Location of phone in pocket\label{fig:lr_pocket_res}]
%{\includegraphics[scale=0.33]{figures/lr_pocket_res.pdf}}
%\subfigure[Detecting the first arriving signal\label{fig:sequence_enter}]
%{\includegraphics[scale=0.33]{figures/sequence_enter.pdf}}
%\vspace{-0.1in}
%\caption{Detecting entering vehicles.}
%\end{figure*}

%As described in the previous section, the location of smartphone will
%effect the observation on the sensory data, thus the system should
%have learnt the activity pattern under such considerations.
%We collect the information of entering vehicle for four cases:
%entering from driver side with smartphone in left pocket, driver side
%with smartphone in right pocket, passenger side with smartphone in
%left pocket, and passenger side with smartphone in right pocket.
%And the test number of each case is $50$.
%The window size is ranging from $1.5s$ to $5s$, and obviously the
%performance is elevating with the increase of window size, as shown in
%Figure~\ref{fig:lr_pocket_res}.
%The accuracy of detection of location of smartphone is about $85\%$
%when the window size is set around $4s$, and also with high
%precision.

Figure~\ref{fig:sequence_enter} shows an illustration of the first
 signal the system detected according to the protocol.
The evaluation is based on the acceleration from the perspective of
the earth, with two dimensions, the horizontal plane and ground.
The whole serials of activities starts from putting the smartphone in
the pocket after making a phone call, and walking towards the door
followed by entering.
As shown in the figure, the system successfully detects the walking
pattern starting from the $112$ time stamp, and after nearly $2$
seconds, the system senses the first arriving signal of entering
($133$ sampling point).
In this evaluation, the window size is $4.5s$, and since the duration
of the entering will last approximately $5$ to $6$ seconds
individually, we slide the window with step length $1s$.
In this case, the system will detect multiple entering behavior, which
we will conclude with high probability that the user is entering the
vehicle.

%\begin{figure*}[!ht]
%\centering
%\subfigure[Accuracy in Left vs. Right\label{fig:win_lr}]{\includegraphics[scale = 0.33]{figures/win_lr.pdf}}
%\subfigure[Side Detection\label{fig:side_res}]{\includegraphics[scale = 0.33]{figures/side_res.pdf}}
%\subfigure[New Training Data\label{fig:side_new_train}]{\includegraphics[scale = 0.33]{figures/side_new_train.pdf}}
%\vspace{-0.1in}
%\caption{Side detection accuracy in different window sizes.}
%\label{fig:win_side}
%\vspace{-0.2in}
%\end{figure*}
After the behavior is determined, the detection of entering side is
followed.
We first evaluate the influence brought about by the window size in
Figure~\ref{fig:win_side}, ranging from $1.5s$ to $5s$ in both
learning and testing.
For both driver side entering and passenger side entering, the
accuracy climbs with the increase of window size, and the accuracy for
both left side and right side are around $90\%$, but the accuracy with
the window size only $1.5s$ is rather low (Figure~\ref{fig:win_lr}).
Both left and right cases have acceptable accuracy when the window
size is around $3s$ with the accuracy over $95\%$.
Figure~\ref{fig:side_res} presents the precision, recall, specificity
 and accuracy for whole process of side detection.
The precision reaches $90\%$ when the window size is $3s$,  the
result still increases while enlarging the window size, and the highest
precision is around $95\%$ with window size $4.5s$.
The total accuracy is approximately $85\%$ when the window size is set
as the largest.
We also evaluate the performance of self-adjusting ability for \ourprotocol
 by introducing new data from another user in Figure~\ref{fig:side_new_train}.
Originally the training data is coming from user $1$, however, such data cannot
 provide convincing results when detecting the data generated from user $2$.
However, with the number of new training data increase, \ourprotocol adjusts automatically,
 and obtains high accuracy, precision and specificity.

\subsection{Front vs. Back}
\begin{figure*}
\vspace{-0.1in}
  \begin{minipage}[t]{0.75\linewidth}
    \centering
    \subfigure[Spike if engine starts.\label{fig:mag_engine_start}]
        {\includegraphics[scale=0.25]{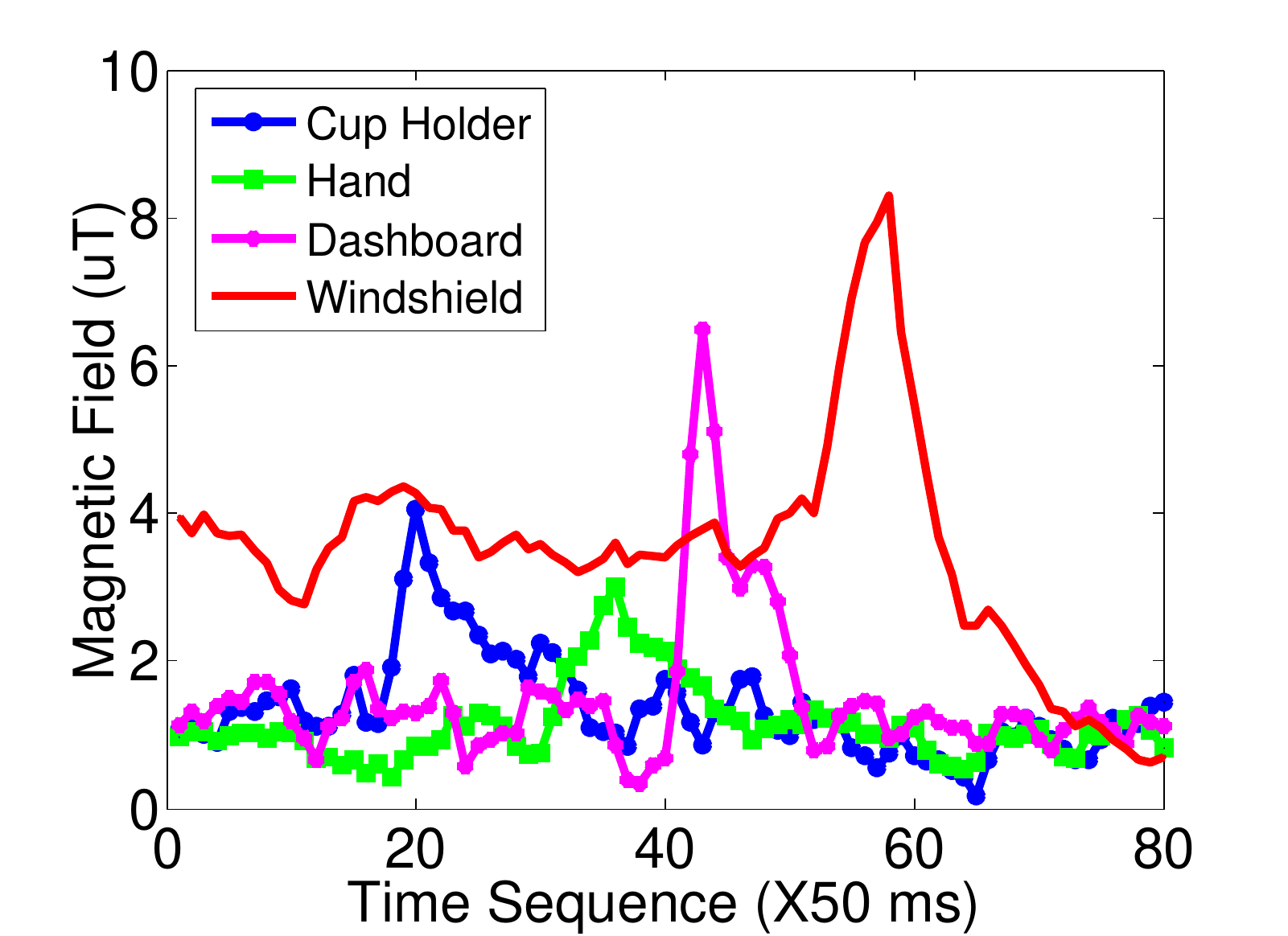}}
    \subfigure[Spatial variations when idle.\label{fig:mag_cars}]
        {\includegraphics[scale=0.25]{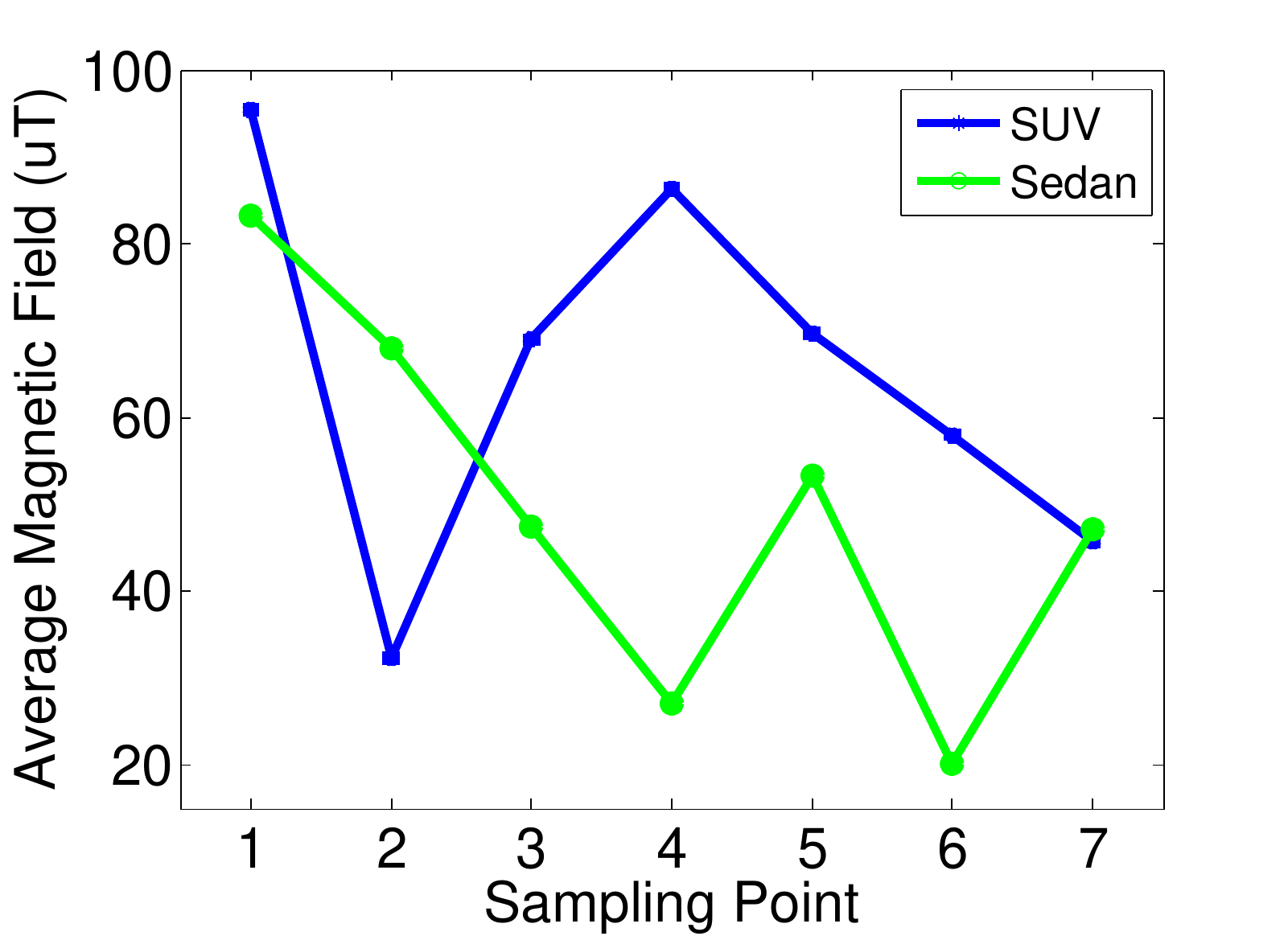}}
    \subfigure[Engine starts.\label{fig:spike}]
        {\includegraphics[scale=0.25]{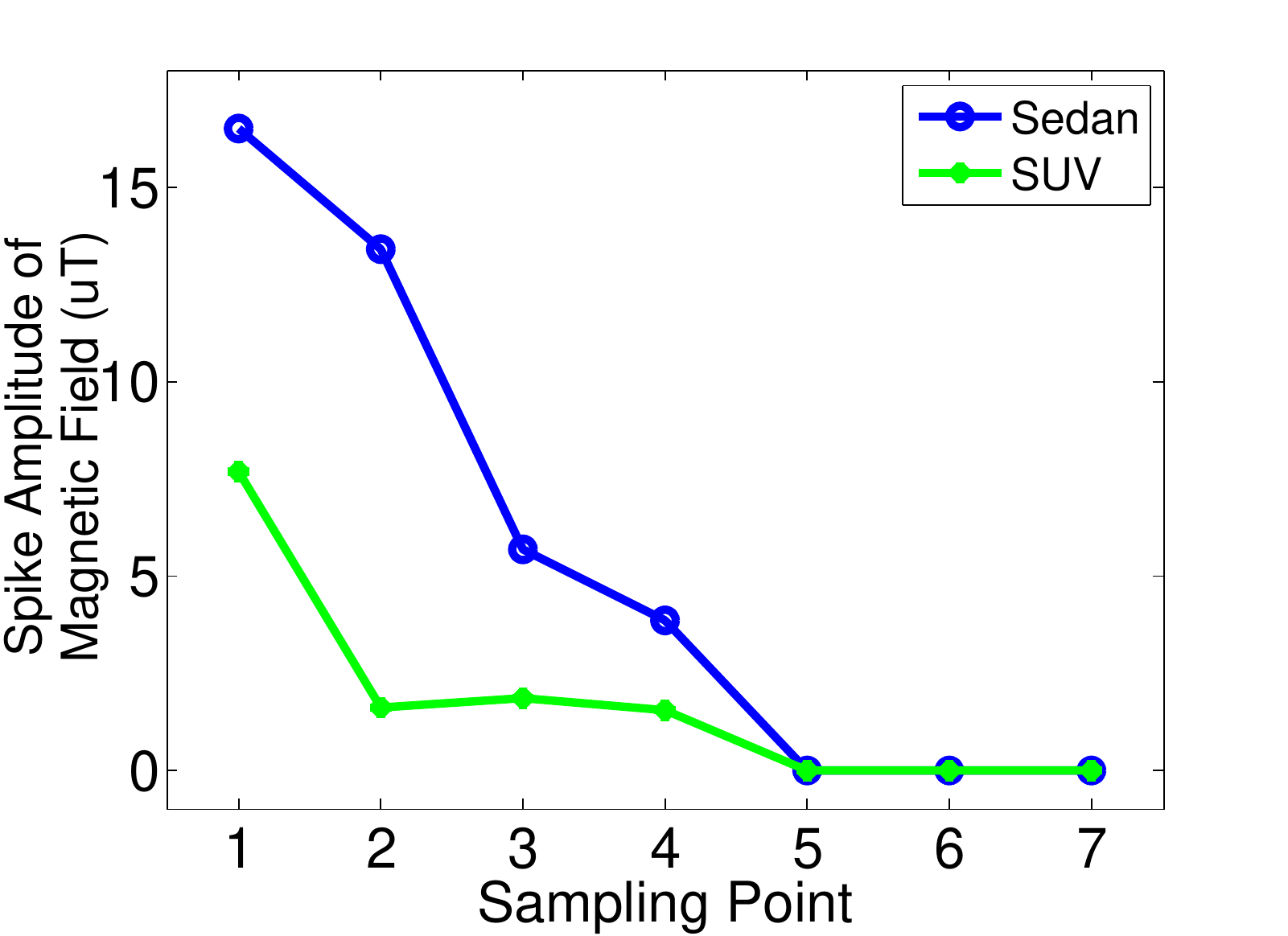}}
    \vspace{-0.1in}
    \caption{Magnetic field fluctuations experienced at different places of car.}
    \vspace{-0.2in}
  \end{minipage}%
  \begin{minipage}[t]{0.2\linewidth}
    \centering
    \includegraphics[scale=0.25]{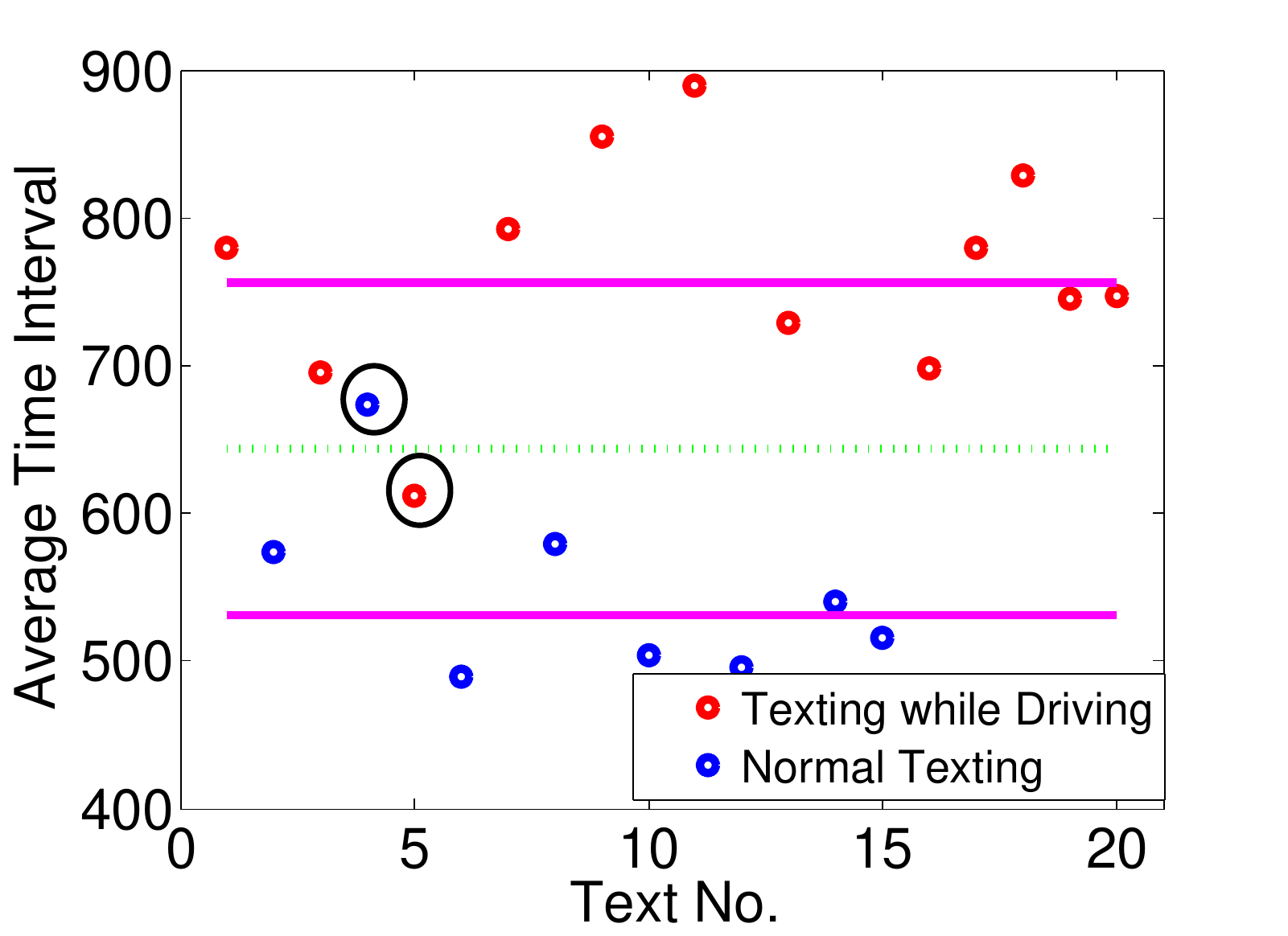}
    \vspace{-0.1in}
        \caption{Texting}
        \label{fig:eva_text2}
    \vspace{-0.2in}
  \end{minipage}
\end{figure*}
%\begin{figure*}[htb]
%\centering
%\subfigure[Spike if engine starts.
%\label{fig:mag_engine_start}]
%{\includegraphics[scale=0.33]{figures/mag_engine_start.pdf}}
%\subfigure[Spatial variations when idle.\label{fig:mag_cars}]{\includegraphics[scale = 0.33]{figures/mag_cars.pdf}}
%\subfigure[Spatial fluctuations if engine starts.\label{fig:spike}]{\includegraphics[scale = 0.33]{figures/spike.pdf}}
%\vspace{-0.1in}
%\caption{Magnetic field fluctuations experienced at different places
%  of car.}
%%\label{fig:mag_in_car}
%\vspace{-0.2in}
%\end{figure*}
Our system presents two independent approaches to handle the
front-back classification through engine start signal monitoring and
bump road signal detecting.
In order to demonstrate the generality of both methods, we organize
the experiment in two different cars (a Mazda sedan and a Honda SUV)
and multiple positions in the cars where the phone may be put.

As mentioned in the previous section, the embedded magnetometer in
smartphone could detect the changing magnetic field when the phone is
located in the trouser pocket.
However, some users get used to make a phone call or texting
while entering the car, and then put in cup holer of under dashboard.
Thus our experiments mainly focus on the detection of the engine start
signal while the smartphone is held in hand or put in some other possible positions
in the car.%
%While the phone is being held in hand, the magnetic field senses a
%small fluctuation on the value because the hand cannot be absolutely
%still.
%Thus we compare the magnetic field for phone in the pocket, phone in
%the hand, and phone in the cup-holder, where three locations are with
%similar distance to the front engine.
%We use two types of smartphones in the test.
%Although the collected level of magnetic field values deviate from each
%other cases in a large gap because of the inherent difference of sensors
%and the random orientation and location, the slight spike of changing
%value can still be detected successfully.

%\begin{figure*}[htb]
%\centering
%\subfigure[Spike if engine starts.
%\label{fig:mag_engine_start}]
%{\includegraphics[scale=0.33]{figures/mag_engine_start.pdf}}
%\subfigure[Spatial variations when idle.\label{fig:mag_cars}]{\includegraphics[scale = 0.33]{figures/mag_cars.pdf}}
%\subfigure[Spatial fluctuations if engine starts.\label{fig:spike}]{\includegraphics[scale = 0.33]{figures/spike.pdf}}
%\vspace{-0.1in}
%\caption{Magnetic field fluctuations experienced at different places
%  of car.}
%\label{fig:mag_in_car}
%\vspace{-0.2in}
%\end{figure*}

In Figure~\ref{fig:mag_engine_start}, we plot the magnetic field
changing when the engine starts in four different locations: cup
holder, holding in hand, in dashboard and under the windshield (sorted
by the distance to the engine).
Obviously, the place closest to the engine experiences largest
fluctuation in the magnetic varying with the amplitude about $7uT$,
with the distance to the engine increases, the amplitude of the
magnetic fluctuation decreases slightly.
When the smartphone is held in the hand while sitting or put in the
cup holder, the amplitude is only half of the value in the
windshield.
We also calculate the variance for the magnetic field value in two
different conditions, and the value is around $0.0614$, $0.0485$ and
$0.0642$ respectively for in hand, pocket and cup-holder when the
engine is off, and approximately $0.3919$, $0.32$ and $0.4860$ when
the engine starts.
Thus, although the magnetic field in the vehicle fluctuates along with
the unpredictable motion behavior of the human body, the orientation,
position and location of the smartphone, the magnetic field can be
considered as a feasible factor to distinguish the front and back.

 Figure~\ref{fig:mag_cars} shows the magnetic field value in both
 Mazda sedan and Honda SUV in seven separate sampling locations, and
 the location numbers indicate the location to the engine in order of increasing
 distance, \ie, under the windshield, dashboard,
 the trouser pocket of driver, in cup-holder,
 back of front seat, back seat, and under the back windshield.
Based on the experiment, the value of magnetic field is determined by
 both the location and position of the smartphone, as well as the
 placement in the vehicle.
The magnetic field is demonstrated to be the largest in both cars
 under the windshield, and decrease when being put close to the
 dashboard, where drivers may put their smartphone while driving.
Although the readings, as shown, are irregular, we still observed
 instant spikes at that very moment, as shown in
Figure~\ref{fig:spike}.
The figure indicates that the closer to the engine, the more sensitive
the magnetic field  variation be, and when put the smartphone in the back
seat area, the sensor can hardly detect the changing magnetic field
when engine starts, which demonstrates that the spike from engine is
 trustable.

We then take two separate sets of experiments in both parking lot and local
roads to evaluate the efficiency of front-back distinguish using bumps
and potholes.
There are one deceleration strip and one bump in the parking lot, and
we drive through both in ten times in each with different driving
speeds.
The test results are shown in Table~\ref{table:res_bump}, both
detections lead to the absolute correctness, $20$ bump are all
successfully detected in both locations.
\begin{table}[hptb]
\vspace{-0.2in}
\caption{Bump on the Road}
\centering
\begin{tabular}{l|l|l}
\hline
   & Bump in Front & Bump in Back\\
\hline
  Test in Front & 20 & 0\\
  Test in Back & 0 & 20\\
\hline
\end{tabular}
\vspace{-0.1in}
%\caption{Bump on the Road}
\label{table:res_bump}
%\vspace{-0.2in}
\end{table}

When it comes to the street test, the results are slightly different.
The experiment is taken in a  suburb at night, the total distance is
approximately $5.2$ miles with local road and highway.
Both the driver and back seat passenger turn on the system to estimate
 its exact location in the car according to the sensory data while
 driving.
The smartphone of driver detects $334$ samples of readings and $23$ of
 bumps and potholes, while the back seat passenger only detects $286$ samples but
 $58$ bumps and potholes.
The sampling number is different because of the starting time of
 passenger is behind the driver.
In addition, although the number of bumps and potholes being detected by both
 smartphones are different, both smartphones report they are in the right
 location with accuracy of $100\%$.

\subsection{Texting Evaluation}

%\begin{figure}[hptb]
%\centering
%%\includegraphics[scale=0.4]{figures/eva_text.pdf}
%\includegraphics[scale=0.4]{figures/eva_text2.pdf}
%\vspace{-0.15in}
%\caption{Texting Evaluation}
%\label{fig:eva_text}
%\vspace{-0.15in}
%\end{figure}
We then  detect the regulation of texting to detect if the user is
driving or not.
We sample $20$ different cases with $8$ texting in normal condition and
 the rest in driving condition tested in the parking lot.
Each sentence is approximately $20$ to $30$ words, we collect the
input time interval and calculate the average value in real time.
In Figure~\ref{fig:eva_text2}, we draw two pink lines,
identifying the average time interval of each  scenario, and the green
dash line in the middle as the standard classifier.
All the dots should be separated by the standard classifier, with the blue (normal texting) below and red (texting while driving) in above.
The two error classification are denoted in black circle.
The evaluation in texting detecting is reliable and feasible, the accuracy is $90\%$.

\subsection{Driver Detection}
The decision of driver detection is based on previous sub-processes
through evidence fusion.
When doing real time recognition, the system slides the window with
step $0.5s$ to match the stored behavior through naive Bayes
classifier.
Since the activity could be detected in multiple times because of the
sliding window, we consider a continuous same activity recognition to
be a successful recognition.
And taking the acceleration into account as a filter, the
recognition could provide high level of credit for current
 recognition.
\CUTIT{%%%%%%%%%%%%%%%%
\begin{figure}[hptb]
\centering
\includegraphics[scale=0.4]{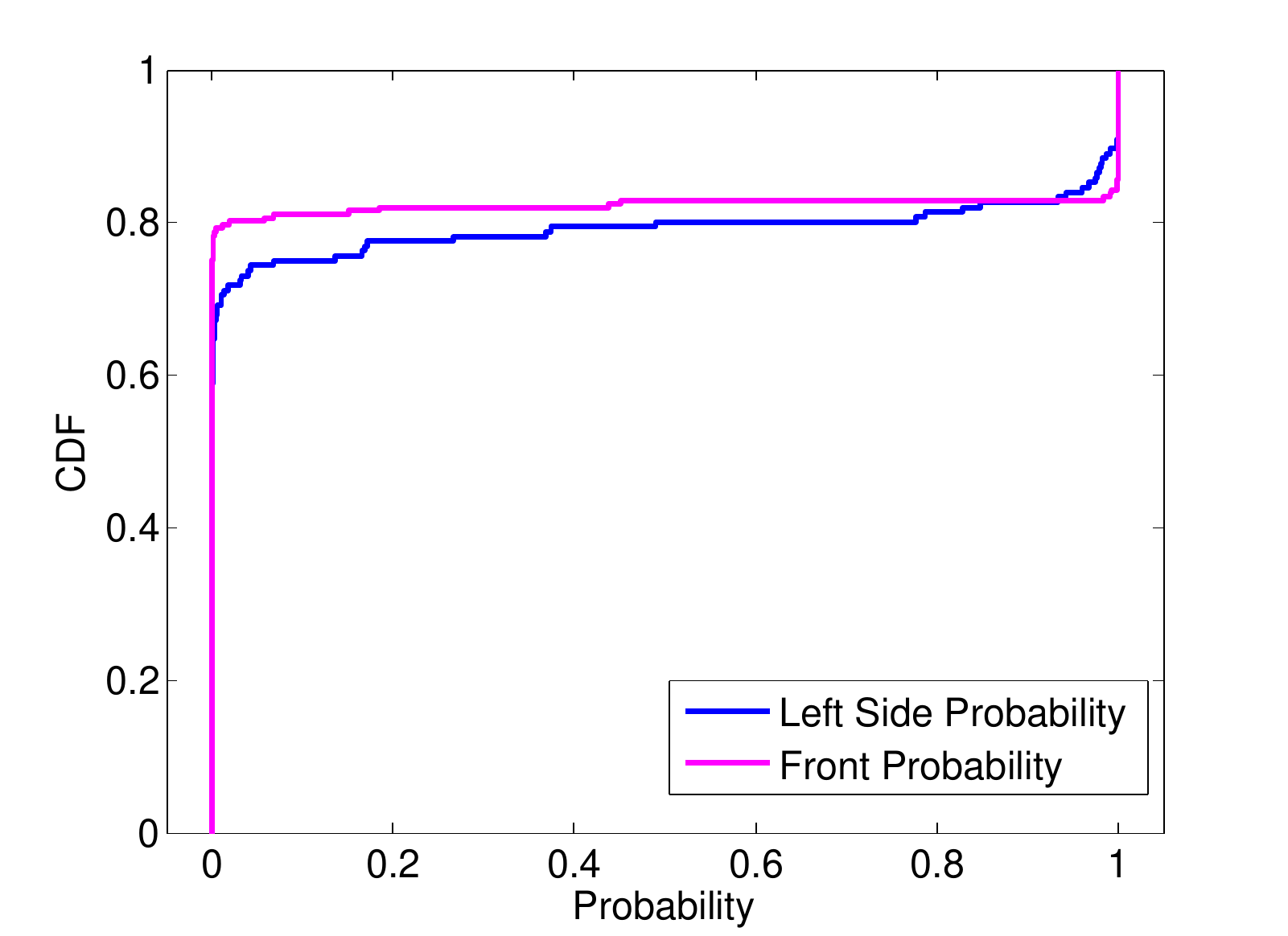}
\vspace{-0.1in}
\caption{bpa of Side and Front}
\label{fig:fusion1}
\end{figure}
}%%%%%%%%%%%%%%%%
%The bpa of both left side detection and front seat is shown in
%Figure~\ref{fig:fusion1}.

%However, since the pattern of different activities are quite
%different, when taking the acceleration into account as a filter,
%the credibility is extremely high.
%We test different locations in the car to calculate the average
%probability of driver when sitting in the exact location.
Based on our experiment, we notice that the performance of \ourprotocol
mainly depends on the first two phases.
We test the performance of driver detection based on the fusion
of all the phases, the precision is $96.67\%$ and accuracy $87.18\%$.
Meanwhile, according to the real evaluation in Android smartphone,
 the recognition delay is only $0.2184$ second.

\subsection{Energy Consumption}
The energy consumption of the system is determined by the running
duration of inertial sensors.
Besides, the Android API provides four different level of sampling
rates, with the energy consumption being largest
($95mA$~\cite{youssef2010gac}) in the fastest level, which the sensory
data being delivered without any delay.
The working strategy of the system is determined based on individual
life pattern, more specifically, the behavior regulation.

We take a group of experiments using Galaxy S$3$ to test the energy
consumption in high density sampling.
Without using any inertial sensors, the battery drop $2\%$ within half
an hour, but $9\%$ when the inertial sensors are triggered.
However, in this process, we reduce the detecting rate to $10s$ in
every one minute with the sensor sampling rate $0.05s$, which on the
other hand, match the transition probability of transferring from
walking to entering car.
Based on the test, the battery reduce only $4\%$ for half an hour.
Other existing works utilize GPS to determine whether the user is in
driving vehicle.
Although such solutions do not require sensors to monitor the behavior
and adjust the user habit, the energy consumption from GPS is much
larger than sensors.
In addition, the system has to open GPS and store sensory data for a
certain duration, depending on when the driving behavior is detected.
In our experiment, the battery discharge from $84\%$ to $70\%$ for the
same testing duration.

\section{Conclusion}
This paper presented \ourprotocol, a smartphone based application to
detect driver and texting  according to user's behavior and
activity pattern.
Our system leverages inertial sensors integrated in smartphone and
accomplish the objective of driver-passenger distinguishing without
relying any additional equipment.
We evaluate each process of the detection, including activity
recognition and show that our system achieves good
sensitivity, specificity, accuracy and precision, which leads to the
high classification accuracy.
Through evaluation, the accuracy of successful detection is
approximately $87.18\%$, and the precision is $96.67\%$.
The evaluation of \ourprotocol is based on the assumption
that smartphone is attached to the user body in the trouser pocket
most of the time.
However, a certain number of users may enter the vehicle while making
a phone call or with their smartphones in the hand bag, which in most
of the time the activities mentioned above may not be detected
precisely sometimes.
Although such conditions may bring us a lot difficulties, the system
is still demonstrated to be robust in handling the detection through
evidence fusion and some side signals.

%%%%%%%%%%%%%%%%%%%%%%%%%%%%%%%%%
%\small{
%\bibliographystyle{acm}
%\bibliography{reference}
%}

\small{

}

\end{document}